\documentclass[a4paper,12pt]{article}
\begin{document}
\begin{titlepage}
\title{\bf
Spectral Properties of the Generalized Spin-Fermion
Models\thanks{International Journal of Modern Physics B13, N 20,
(1999) 2573-2605} }
\author{
A.L.Kuzemsky \thanks
{E-mail:kuzemsky@thsun1.jinr.ru ;
http://thsun1.jinr.ru/~ kuzemsky}
\\
Bogoliubov Laboratory of Theoretical Physics, \\
Joint Institute for Nuclear Research,\\
141980 Dubna, Moscow Region, Russia.}
\date{}
\maketitle
\begin{abstract}
In order to account for competition and interplay of localized and
itinerant magnetic behaviour in correlated many body systems with
complex spectra  the various types of  spin-fermion models have
been considered in the context of the Irreducible Green's
Functions (IGF) approach. Examples are generalized  $d-f$ model
and Kondo-Heisenberg model. The  calculations of the quasiparticle
excitation spectra with damping for these models has been
performed in the framework of the equation- of-motion method for
two-time temperature Green's Functions within a non-perturbative
approach. A unified scheme for the construction of Generalized
Mean Fields (elastic scattering corrections) and self-energy
(inelastic scattering) in terms of the Dyson equation has been
generalized in order to include the presence of the two
interacting subsystems of localized spins and itinerant electrons.
A general procedure is given to obtain the quasiparticle damping
in a self-consistent way. This approach gives the complete and
compact description of quasiparticles and show the flexibility and
richness of the generalized spin-fermion model concept.
\end{abstract}
\end{titlepage} \newpage
\section{Introduction}
The existence and properties of localized and itinerant magnetism
in metals, oxides and alloys and their interplay is an interesting
and not yet fully understood problem of quantum theory of
magnetism.  The behaviour and the true nature of the electronic
and spin states and their quasiparticle dynamics are of central
importance to the understanding of the physics of correlated
systems such as magnetism and Mott-Hubbard metal-insulator
transition in metal and oxides, magnetism and heavy fermions (HF)
in rare-earths compounds, high-temperature superconductivity
(HTSC) in cuprates and anomalous transport properties in
perovskite manganates.  This class of systems are characterized by
the complex, many-branch spectra of elementary excitations and,
moreover, the correlations effects are essential.\\ Recently there
has been considerable interest in identifying the microscopic
origin of quasiparticle states in such systems and a number of
model approaches have been proposed.  A principle importance  of
these studies is concerned with a fundamental problem of
electronic solid state theory, namely with the tendency of 3d
electrons in transition metal compounds and 4f electrons in
rare-earth metal compounds  to exhibit both localized and
delocalized behaviour.  The interesting electronic and magnetic
properties of these substances are intimately related to this dual
behaviour of electrons.  In spite of experimental and theoretical
achievements still it remains much to be understood concerning
such
systems.\\
Many magnetic and electronic properties of these materials may be
interpreted in terms of combined
spin-fermion models (SFM), which include the interacting spin and charge
subsystems.
This add the richness to physical behaviour and brings in
significant and interesting physics, e.g. the bound states and magnetic
polarons, HF, and colossal negative magnetoresistance.\\
The problem of the adequate description of quasiparticle many-body
dynamics of generalized spin-fermion models has been studied
intensively during the last decades, especially in the context of
magnetic and transport properies of rare-earth and transition
metals and their compounds~\cite{coq1} - \cite{imad}, magnetic
semiconductors~\cite{kuz4},~\cite{kuz5}, interplay of magnetism
and HF ~\cite{ cole},\cite{coq6} , HTSC ~\cite{pre7}
-~\cite{kuz12} and magnetic and transport properties of perovskite
manganates~\cite{dege13}, \cite{mae14}, \cite{imad}. Variety of
metal-insulator transitions and correlated metals phenomena in
$d(f)$-electron systems as well as the relevant models have been
comprehensively discussed recently in Ref.~\cite{imad}.\\ The
basic theory of the physical behaviour of SFMs has been studied
mainly within mean field approximation.  However many experimental
investigations call for a better understanding of the nature of
solutions (especially magnetic) to the spin-fermion and related
correlated models, such as $t-J$, Kondo-Heisenberg model,
etc~\cite{kuz10},\cite{imad}. \\ In the previous papers we set up
the formalism of the method of the Irreducible Green's Functions
(IGF)~\cite{kuz15} -\cite {kuz23}.  This IGF method allows one to
describe the quasiparticle spectra with damping for the systems
with complex spectra and strong correlation in a very general and
natural way. This scheme differs from the traditional method of
decoupling of the infinite chain of the equations~\cite{tyab24}
and permits to construct the relevant dynamical solutions in a
self-consistent way on the level of the Dyson equation without
decoupling the chain of the equation of motion for the GFs.\\ In
this paper we apply the formalism to consider the quasiparticle
spectra for the complex systems, consisting of a few interacting
subsystems.  It is the purpose of this paper to explore more fully
the notion of Generalized Mean Fields (GMF)~\cite{kuz10} which may
arise in the system of interacting localized spins and lattice
fermions to justify and understand the "nature" of relevant
mean-fields and damping effects.\\ It is worthy to emphasize that
in order to understand quantitatively the electrical, thermal and
superconducting properties of metals and their alloys one needs a
proper description an electron-lattice interaction
too~\cite{kuz25}-~\cite{kuz27}.  A systematic, self-consistent
simultaneous treatment of the electron-electron and
electron-phonon interaction plays an important role in recent
studies of strongly correlated systems~\cite{kuz22}.  The natural
approach for the description of electron-lattice effects in such
type of compounds is the Modified Tight-Binding Approximation
(MTBA)~\cite{kuz26}, \cite{kuz27}. We shall consider here the
effects of electron-lattice interaction within the spin-fermion
model approach.\\ The purpose of this paper is to extend the
general analysis to obtain the quasiparticle spectra and their
damping of the concrete model systems consisting of two or more
interacting subsystems within various types of spin-fermion models
to extend their applicability and show the effectiveness of IGF
method.
\section{Irreducible Green's Functions Method}
In this paper we will use the IGF approach which allows one to
describe completely the quasi-particle spectra with damping in a very
natural way.
The essence of our consideration of the dynamical properties of
many-body system with complex spectra and strong interaction is related
closely with the field theoretical approach and use the advantage of
the Green's functions language and the Dyson equation. It is possible
to say that our method tend to emphasize the fundamental and central
role of the Dyson equation for the single-particle dynamics of the
many-body systems at finite temperatures.\\ In this Section, we will
discuss briefly this novel nonperturbative approach for the description
of the many-body dynamics of many branch and strongly correlated
systems.  The considerable progress in studying the spectra of
elementary excitations and thermodynamic properties of many-body
systems has been for most part due to the development of the
temperature dependent Green's Functions methods. We have developed a
helpful reformulation of the two-time GFs method~\cite{tyab24} which is
especially adjusted~\cite{kuz15} for the correlated fermion systems on
a lattice and systems with complex spectra~\cite{kuz4},\cite{kuz5}. The
similar method has been proposed in Ref.~\cite{plak} for Bose
systems( anharmonic phonons and spin dynamics of pure Heisenberg
ferromagnet). The very important concept of the whole method are the
{\bf Generalized Mean Fields}.  These GMFs have a complicated structure
for the strongly correlated case  and complex spectra and do not reduce
to the functional of the mean densities of the electrons or spins, when
we calculate excitations spectra at finite temperatures. \\ To clarify
the foregoing, let us consider the retarded GF of the
form~\cite{tyab24} \begin{equation} G^{r} = <<A(t), B(t')>> =
-i\theta(t - t')<[A(t)B(t')]_{\eta}>, \eta = \pm 1.  \end{equation} As
an introduction of the concept of IGFs let us describe the main ideas
of this approach in a symbolic form. To calculate the retarded GF $G(t
- t')$ let us write down the equation of motion for it:
\begin{equation} \omega G(\omega) = <[A, A^{+}]_{\eta}> + <<[A,
H]_{-}\mid A^{+}>>_{\omega}.  \end{equation} The essence of the method
is as follows~\cite{kuz18}. It is based on the notion of the {\it
``IRREDUCIBLE"} parts of GFs (or the irreducible parts of the
operators, out of which the GF is constructed) in term of which it is
possible, without recourse to a truncation of the hierarchy of equations
for the GFs, to write down the exact Dyson equation and to obtain an
exact analytical representation for the self-energy operator. By definition
we introduce the irreducible part {\bf (ir)} of the GF
\begin{equation}
^{ir}<<[A, H]_{-}\vert A^{+}>> = <<[A, H]_{-} - zA\vert A^{+}>>.
\end{equation}
The unknown constant z is defined by the condition (or constraint)
\begin{equation}
<[[A, H]^{ir}_{-}, A^{+}]_{\eta}> = 0
\end{equation}
From the condition (4) one can find:
\begin{equation}
z = \frac{<[[A, H]_{-}, A^{+}]_{\eta}>}{<[A, A^{+}]_{\eta}>} =
 \frac{M_{1}}{M_{0}}
\end{equation}
Here $M_{0}$ and $M_{1}$ are the zeroth and first order moments of the
spectral density. Therefore, irreducible GF  are defined so that they
cannot be reduced to the lower-order ones by any kind of decoupling. It
is worthy to note that the irreducible correlation functions are well
known in statistical mechanics. In the diagrammatic approach the
irreducible vertices are defined as the graphs that do not contain
inner parts connected by the $G^{0}$-line. With the aid of the definition
(3) these concepts are translated into the language of retarded and
advanced GFs. This procedure extract all relevant (for the problem under
consideration) mean field contributions and puts them into the generalized
mean-field GF, which here are defined as
\begin{equation}
G^{0}(\omega) = \frac{<[A, A^{+}]_{\eta}>}{(\omega - z)}.
\end{equation}
To calculate the IGF $ ^{ir}<<[A, H]_{-}(t), A^{+}(t')>>$ in (2), we
have to write the equation of motion after differentiation with respect
to the second time variable $t'$. The condition (4) removes the
inhomogeneous term from this equation and is a very crucial point of
the whole approach. If one introduces an irreducible part for the
right-hand side operator as discussed above for the ``left" operator,
the equation of motion (2) can be exactly rewritten in the following
form \begin{equation} G = G^{0} + G^{0}PG^{0}.  \end{equation} The
scattering operator $P$ is given by \begin{equation} P =
(M_{0})^{-1}\quad ^{ir}<<[A, H]_{-}\vert[A^{+}, H]_{-}>>^{ir}
(M_{0})^{-1}.
\end{equation}
The structure of the equation (7) enables us to determine the
self-energy operator $M$, in complete analogy with the diagram
technique \begin{equation} P = M + MG^{0}P.  \end{equation} From the
definition (9) it follows that  the self-energy
operator $M$ is defined as a proper (in diagrammatic language
``connected") part of the scattering operator $M = (P)^{p}$. As a
result, we obtain the exact Dyson equation for the thermodynamic
two-time Green's Functions:  \begin{equation} G = G^{0} + G^{0}MG,
\end{equation}
which has a well known formal solution of the form
\begin{equation}
G = [ (G^{0})^{-1} - M ]^{-1}; \quad  M =  G^{-1}_{0}  - G^{-1}
\end{equation}
Thus, by introducing irreducible parts of GF (or the irreducible
parts of the operators, out of which the GF is constructed) the
equation of motion (2) for the GF can be exactly (but using
constraint (4)) transformed into Dyson equation for the two-time
thermal GF. This is very remarkable result, which deserves
underlining, because of the traditional form of the GF method did
not include  this point. The projection operator technique has
essentially the same philosophy, but with using the constraint (4)
in our approach we emphasize the fundamental and central role of
the Dyson equation for the calculation of the single-particle
properties of the many-body systems. It is important to note, that
for the retarded and advanced GFs the notion of the proper part is
symbolic in nature~\cite{kuz18}. However, because of the identical
form of the equations for the GFs for all three types (advanced,
retarded and causal), we can convert in each stage of calculations
to causal GFs and, thereby, confirm the substantiated nature of
definition (9)! We therefore should speak of an analogue of the
Dyson equation. Hereafter we will drop this stipulation, since it
will not cause any misunderstanding. It should be emphasized that
the scheme presented above give just an general idea of the IGF
method. The specific method of introducing IGFs depends on the
form of operator $A$, the type of the Hamiltonian and the
conditions of the problem. The general philosophy of the IGF
method lies in the separation and identification of elastic
scattering effects and inelastic ones. This last point is quite
often underestimated and both effects are mixed. However, as far
as the right definition of quasiparticle damping is concerned, the
separation of elastic and inelastic scattering processes is
believed to be crucially important for the many-body systems with
complicated spectra and strong interaction.   From a technical
point of view the elastic (GMF) renormalizations can exhibit a
quite non-trivial structure. To obtain this structure correctly,
one must construct the full GF from the complete algebra of the
relevant operators and develop a special projection procedure for
higher-order GF in accordance with a given algebra. \\ It is
necessary to emphasize that there is an intimate connection
between the adequate introduction of mean fields and internal
symmetries of the Hamiltonian. In many-body interacting systems,
the symmetry is important in classifying of the different phases
and in understanding of the phase transitions between them. The
problem of finding of the ferromagnetic and antiferromagnetic
"symmetry broken" solutions of the correlated lattice fermion
models within IGF method has been investigated in
Ref.~\cite{kuz23}. A unified scheme for the construction of
Generalized Mean Fields (elastic scattering corrections) and
self-energy (inelastic scattering) in terms of the Dyson equation
has been generalized in order to include the presence of the
"source fields". The "symmetry broken" dynamical solutions of the
Hubbard model, which correspond to various types of itinerant
antiferromagnetism has been discussed. This approach complements
previous studies of microscopic theory of Heisenberg
antiferromagnet~\cite{kuz17} and clarifies the nature of the
concepts of Neel sublattices for localized and itinerant
antiferromagnetism and "spin-aligning fields" of correlated
lattice fermions.
\section{Quasiparticle Dynamics of the $d-f$ Model}
\subsection{Generalized $d-f$ model}
The concept of the $s(d)-f$ model play an important role in the
quantum theory of magnetism~\cite{coq1},\cite{kuz25}. In this
section we consider the generalized $d-f$ model, which describe
the localized $4f(5f)$-spins interacting with $d$-like
tight-binding itinerant electrons and take into consideration the
electron-electron and electron-phonon interaction in the framework
of MTBA ~\cite{kuz26},\cite{kuz27}.\\ The total Hamiltonian of the
model is given by \begin{equation} H = H_{d} + H_{d-f} + H_{d-ph}
+ H_{ph}
\end{equation} The Hamiltonian of tight-binding electrons is given by
\begin{equation} H_{d} = \sum_{ij} \sum_{\sigma}
t_{ij}a^{+}_{i\sigma}a_{j\sigma} + \frac{1}{2} \sum_{i\sigma}
Un_{i\sigma}n_{i-\sigma} \end{equation} This is the Hubbard
model~\cite{hub28}.  The band energy of Bloch electrons $\epsilon(\vec
k)$ is defined as follows $$t_{ij} = N^{-1}\sum_{\vec k}\epsilon(\vec
k) \exp[i{\vec k}({\vec R_{i}} -{\vec R_{j}})],$$ where  $N$ is the
number of the lattice sites.  For the tight-binding electrons in cubic
lattice we use the standard expression for the dispersion
\begin{equation} \epsilon(\vec k) = 2\sum_{\alpha}t( \vec
a_{\alpha})\cos(\vec k \vec a_{\alpha}) \end{equation}, where $\vec
a_{\alpha}$ denotes the lattice vectors in a simple lattice with
inversion centre.\\ The term $H_{d-f}$ describes the interaction of the
total 4f(5f)-spins with the spin density of the itinerant electrons
\begin{equation} H_{d-f} = \sum_{i}J{\vec \sigma_{i}}{\vec S_{i}} = - J
N^{-1/2}\sum_{kq}\sum_{\sigma}[S^{-\sigma}_{-q}a^{+}_{k\sigma}
a_{k+q-\sigma} + z_{\sigma}S^{z}_{-q}a^{+}_{k\sigma}a_{k+q\sigma}]
\end{equation}
where
sign factor $z_{\sigma}$ is given by $$z_{\sigma} = (+ or -)\quad
for \quad \sigma =  (\uparrow  or  \downarrow)$$ and
$$S^{-\sigma}_{-q} = \cases {S^{-}_{-q} &if $\sigma = +$ \cr
S^{+}_{-q} &if $\sigma = -$ \cr}$$
In general the indirect exchange integral $J$ strongly depends on the
wave vectors $J(\vec k; \vec k+ \vec q)$ having its maximum value at
$k=q=0$.  We omit this dependence for the sake of brevity  of notations
only. \\ For the electron-phonon interaction we use the following
Hamiltonian~\cite{kuz26}
\begin{equation}
H_{d-ph} = \sum_{\nu\sigma}\sum_{kq}
V^{\nu}(\vec k, \vec k + \vec q)Q_{\vec q\nu}a^{+}_{k+q\sigma}
a_{k\sigma}
\end{equation}
where
\begin{equation}
V^{\nu}(\vec k, \vec k + \vec q) =
\frac{2iq_{0}}{( N M )^{1/2}}\sum_{\alpha}
t(\vec
a_{\alpha})e^{\alpha}_{\nu}(\vec q)[\sin \vec a_{\alpha} \vec k
- \sin \vec a_{\alpha} (\vec k - \vec q)]
\end{equation}
here $q_{0}$ is the Slater coefficient~\cite{kuz26} originated in
the exponential decrease of the wave functions of $d$-electrons, N
is the number of unit cells in the crystal and M is the ion mass.
The $\vec
e_{\nu}(\vec q)$ are the polarization vectors of the phonon modes.\\
For the ion subsystem we have
\begin{equation}
H_{ph} =
\frac{1}{2} \sum_{q\nu}
(P^{+}_{q\nu}P_{q\nu} +
\omega^{2}(\vec q \nu)Q^{+}_{q\nu}Q_{q\nu})
\end{equation}
where
$P_{q\nu}$ and $Q_{q\nu}$ are the normal coordinates and
$\omega(q\nu)$ are the acoustical phonon frequencies. Thus, as in
Hubbard model~\cite{hub28}, the $d$- and $s(p)$-bands are replaced by
one effective band in our  $d-f$ model. However, the $s$-electrons give
rise to screening effects and are taken into effects by choosing proper
values of $U$ and $J$ and the acoustical phonon frequencies.
\subsection{Spin Dynamics of the $d-f$ Model}
In this section, to make the discussion more concrete and to illustrate
the nature of spin excitations in the $d-f$ model we  consider
the double-time thermal GF of localized spins~\cite{tyab24}, which is
defined as \begin{eqnarray} G^{+-}(k;t - t') =
<<S^{+}_{k}(t),S^{-}_{-k}(t')>> =
-i\theta(t - t')<[S^{+}_{k}(t),S^{-}_{-k}(t')]_{-}> =
\nonumber\\ 1/2\pi \int_{-\infty}^{+\infty} d\omega \exp(-i\omega t)
G^{+-}(k;\omega) \end{eqnarray}
The next step is to write down the equation of motion for the
GF.
Our attention will be focused on spin dynamics of the model. To
describe self-consistently the spin dynamics of the $d-f$ model one
should take into account the full algebra of relevant operators of the
suitable "spin modes", which are appropriate when the goal is to
describe self-consistently the quasiparticle spectra of two interacting
subsystems. This relevant algebra should be described by the 'spinor'
${\vec S_{i}\choose \vec \sigma_{i}}$ ("relevant degrees of freedom"),
according to IGF strategy of Section 2. \\ Once this has been done,
we must introduce the generalized matrix GF of the form
\begin{equation} \pmatrix{ <<S^{+}_{k}\vert S^{-}_{-k}>> &
<<S^{+}_{k}\vert \sigma^{-}_{-k}>> \cr <<\sigma^{+}_{k}\vert
S^{-}_{-k}>> & <<\sigma^{+}_{k}\vert \sigma^{-}_{-k}>> \cr} = \hat
G(k;\omega) \end{equation} Here $$\sigma^{+}_{k} = \sum_{q}
a^{+}_{k\uparrow}a_{k+q\downarrow} ;\quad \sigma^{-}_{k} = \sum_{q}
a^{+}_{k\downarrow}a_{k+q\uparrow} $$
To explore the advantages of the IGF in the most full form, we shall
do the calculations in the matrix form.\\
To demonstrate the utility of the IGF method we consider the following
steps in a more detail form. Differentiating the GF
$<<S^{+}_{k}(t) \vert B (t')>> $ with respect to the first time, $t$, we
find
\begin{eqnarray}
\omega<<S^{+}_{k} \vert B >>_{\omega} = {2N^{-1/2}<S^{z}_{0}>\brace 0}
+ \\ \nonumber \frac {J}{N} \sum_{pq}
<<2S^{z}_{k-q}a^{+}_{p\uparrow}a_{p+q\downarrow} -
S^{+}_{k-q}(a^{+}_{p\uparrow}a_{p+q\uparrow} -
a^{+}_{p\downarrow}a_{p+q\downarrow}) \vert B>>_{\omega}
\end{eqnarray}
where
$$ B = {S^{-}_{-k} \brace \sigma^{-}_{-k}}
$$\\
Let us introduce by definition irreducible $(ir)$ operators as
\begin{eqnarray}
(S^{z}_{k-q})~^{ir} = S^{z}_{k-q} -<S^{z}_{0}>\delta_{k,q}\\
(a^{+}_{p\uparrow}a_{p+q\uparrow})^{~ir} =
a^{+}_{p\uparrow}a_{p+q\uparrow} -
<a^{+}_{p\uparrow}a_{p\uparrow}>\delta_{q,0}
\nonumber
\end{eqnarray}
Using the definition of the irreducible parts the equation of motion
(21) can be exactly transformed to the following form \begin{eqnarray}
(\omega - JN^{-1}(n_{\uparrow} - n_{\downarrow})) <<S^{+}_{k} \vert B
>>_{\omega} + 2JN^{-1}<S^{z}_{0}><<\sigma^{+}_{k} \vert
S^{-}_{-k}>>_{\omega} = \\ \nonumber
{2N^{-1/2}<S^{z}_{0}>\brace 0} \mp \\ \nonumber \frac
{J}{N} \sum_{pq}
<<2(S^{z}_{k-q})^{~ir}a^{+}_{p\uparrow}a_{p+q\downarrow} -
S^{+}_{k-q}(a^{+}_{p\uparrow}a_{p+q\uparrow} -
a^{+}_{p\downarrow}a_{p+q\downarrow})^{~ir} \vert B>>_{\omega}
\end{eqnarray}
where
$$n_{\sigma} = \sum_{q} <a^{+}_{q\sigma}a_{q\sigma}> = \sum_{q}
f_{q\sigma} =\sum_{q}(\exp(\beta\epsilon(q\sigma)) + 1) $$
To write down the equation of
motion for the Fourier transform of the GF
$<<\sigma^{+}_{k}(t),B(t')>>$
we need the  following auxiliary equation of motion
\begin{eqnarray}
(\omega + \epsilon(p) - \epsilon(p-k) - 2JN^{-1/2}<S^{z}_{0}> -
UN^{-1}(n_{\uparrow} - n_{\downarrow}))
<<a^{+}_{p\uparrow}a_{p+k\downarrow} \vert B >>_{\omega} +\\
\nonumber
UN^{-1}(f_{p\uparrow} - f_{p+k\downarrow}<<\sigma^{+}_{k} \vert
B>>_{\omega} +
JN^{-1/2}(f_{p\uparrow} - f_{p+k\downarrow}<<S^{+}_{k} \vert
B>>_{\omega} =\\
\nonumber
{0 \brace (f_{p\uparrow} - f_{p+k\downarrow})} -
JN^{-1/2} \sum_{qr} <<S^{+}_{-r}(a^{+}_{p\uparrow}a_{q+r\uparrow}
\delta_{p+k,q} -
a^{+}_{q\downarrow}a_{p+k\downarrow}
\delta_{p,q+r})^{~ir}
\vert B>>_{\omega} - \\
\nonumber
JN^{-1/2} \sum_{qr}
<<(S^{z}_{-r})^{~ir}(a^{+}_{q\uparrow}a_{p+k\downarrow}
\delta_{p,q+r} + a^{+}_{p\uparrow}a_{q+r\downarrow}
\delta_{p+k,q}) \vert B>>_{\omega} + \\
\nonumber
UN^{-1} \sum_{qr}
<<(a^{+}_{p\uparrow}a^{+}_{q+r\uparrow}a_{q\uparrow}a_{p+r+k\downarrow}
-
a^{+}_{p+r\uparrow}a^{+}_{q-r\downarrow}a_{q\downarrow}a_{p+k\downarrow})^{~ir}
\vert B>>_{\omega} + \\
\nonumber
\sum_{\nu q}
<<(V^{\nu}(q,k+p-q)a^{+}_{p\uparrow}a_{k+p-q\downarrow} -
V^{\nu}(q,p)a^{+}_{p+q\uparrow}a_{k+p\downarrow})Q_{q\nu}
\vert B>>_{\omega}
\nonumber
\end{eqnarray}
Let us use now the following notations:
\begin{eqnarray}
A = \frac
{J}{N} \sum_{pq}
[2(S^{z}_{k-q})^{~ir}a^{+}_{p\uparrow}a_{p+q\downarrow} -
S^{+}_{k-q}(a^{+}_{p\uparrow}a_{p+q\uparrow} -
a^{+}_{p\downarrow}a_{p+q\downarrow})^{~ir}]; \\ \nonumber
B_{p} = JN^{-1/2} \sum_{qr}
[S^{+}_{-r}(a^{+}_{p\uparrow}a_{q+r\uparrow} \delta_{p+k,q} -
a^{+}_{q\downarrow}a_{p+k\downarrow}
\delta_{p,q+r})^{~ir} - \\
\nonumber
(S^{z}_{-r})^{~ir}(a^{+}_{q\uparrow}a_{p+k\downarrow}
\delta_{p,q+r} + a^{+}_{p\uparrow}a_{q+r\downarrow}
\delta_{p+k,q}) ] + \\
\nonumber
UN^{-1} \sum_{qr}
(a^{+}_{p\uparrow}a^{+}_{q+r\uparrow}a_{q\uparrow}a_{p+r+k\downarrow}
-
a^{+}_{p+r\uparrow}a^{+}_{q-r\downarrow}a_{q\downarrow}a_{p+k\downarrow})^{~ir}
+ \\
\nonumber
\sum_{\nu q}
(V^{\nu}(q,k+p-q)a^{+}_{p\uparrow}a_{k+p-q\downarrow} -
V^{\nu}(q,p)a^{+}_{p+q\uparrow}a_{k+p\downarrow}); \\
\nonumber
\Omega_{1} = \omega - JN^{-1}(n_{\uparrow} - n_{\downarrow}); \quad
\Omega_{2} = 2JN^{-1}<S^{z}_{0}>;\\ \nonumber
\omega_{p,k} = (\omega + \epsilon(p) - \epsilon(p+k) - \Delta);\\
\nonumber
\Delta =
2JN^{-1/2}<S^{z}_{0}> - UN^{-1}(n_{\uparrow} - n_{\downarrow});\\
\nonumber
\chi ^{df}_{0}(k,\omega) = N^{-1} \sum_{p} \frac {
(f_{p+k\downarrow} - f_{p\uparrow})}{\omega_{p,k}};
\end{eqnarray}
In the matrix notations the full equation of motion can be
summarized now in the following form \begin{equation} \hat \Omega \hat
G(k;\omega) = \hat I +  \sum_{p}\hat \Phi(p) \pmatrix{ <<A \vert
S^{-}_{k}>> & <<A \vert \sigma^{-}_{-k}>> \cr <<B_{p} \vert
S^{-}_{-k}>> & <<B_{p} \vert \sigma^{-}_{-k}>> \cr} \end{equation}
where \begin{eqnarray} \hat \Omega = \pmatrix{ \Omega_{1} &
\Omega_{2}\cr -JN^{-1/2}\chi^{df}_{0}&(1 - U\chi^{df}_{0} ) \cr} ;
\quad \hat I = \pmatrix{ J^{-1}N^{1/2}\Omega_{2} & 0\cr
0&-N\chi^{df}_{0}\cr} ; \\ \nonumber
\hat \Phi(p) =
\pmatrix{
N^{-1} & 0\cr
0&\omega^{-1}_{p,k}\cr} ;
\end{eqnarray}
To calculate the higher-order GFs in (26),
we will differentiate the r.h.s. of it with respect to
the second-time variable (t'). Combining both (the
first- and second-time differentiated) equations of
motion we get the "exact"( no approximation have been made till now)
"scattering" equation
\begin{equation}
\hat \Omega \hat G(k;\omega) = \hat I
+  \sum_{pq}\hat \Phi(p) \hat P(p,q) \hat \Phi(q) (\hat
\Omega^{+})^{-1}
\end{equation}
This equation can be identically transformed to the standard form (7)
\begin{equation}
\hat G = \hat G_{0} +
\hat G_{0}
\hat P \hat G_{0}
\end{equation}
Here we have introduced the generalized mean-field (GMF) GF $G_{0}$
according to the following definition
\begin{equation}
\hat G_{0} =
\hat \Omega^{-1} \hat I
\end{equation}
The scattering operator $P$ has the form
\begin{equation}
\hat P =
  \hat I^{-1} \sum_{pq} \hat \Phi(p) \hat P(p,q) \hat \Phi(q) \hat
I^{-1}
\end{equation}
Here we have used the obvious notation
\begin{equation}
\hat P(k,q;\omega) =
\pmatrix{ <<A^{ir} \vert \tilde A^{ir}>> &
<<A^{ir} \vert \tilde B^{ir}_{q}>> \cr <<B_{p}^{ir} \vert
\tilde A^{ir}>> & <<B_{p}^{ir} \vert \tilde B^{ir}_{q}>> \cr}
\end{equation}
The operators $\tilde A$ and $\tilde B_{q}$ follow from $A$ and $B_{q}$
by interchange $\uparrow \rightarrow \downarrow$, $\vec k
\rightarrow -\vec k$ and $S^{+} \rightarrow -S^{-}$.\\
As shown in Section 2, equation (29) can be transformed exactly into a
Dyson equation (10) by means of the definition (9).  Hence, the
determination of the full GF $\hat G$ has been reduced to the
determination of $\hat G_{0}$ and $\hat M$.
\subsection{Generalized Mean-Field GF }
From the definition (30) the GF matrix in generalized mean-field
approximation reads
\begin{equation}
\hat G_{0} = R^{-1}
\pmatrix{
(1 - U\chi^{df}_{0} )
J^{-1}N^{1/2}\Omega_{2}
& \Omega _{2} N \chi^{df}_{0}\cr
\Omega _{2} N \chi^{df}_{0}
&-\Omega_{1}N\chi^{df}_{0}\cr}
\end{equation}
where
$$R =
(1 - U\chi^{df}_{0} )\Omega_{1}
+ \Omega _{2}J N^{1/2} \chi^{df}_{0}$$ \\
The spectrum of quasiparticle excitations without damping
follows from the poles of the generalized mean-field GF (33).\\
Let us write down explicitly the first matrix element $G^{11}_{0}$
\begin{equation}
<<S^{+}_{k} \vert S^{-}_{-k}>>^{0} =
\frac {2JN^{-1/2}<S^{z}_{0}>}{
\omega - JN^{-1}(n_{\uparrow} - n_{\downarrow}) +
2J^{2}N^{-1/2}<S^{z}_{0}>
(1 - U\chi^{df}_{0} )^{-1}\chi^{df}_{0}}
\end{equation}
This result can be considered  as reasonable approximation for
description of the dynamics of localized spins in heavy rare-earth
metals like $Gd$.  (c.f.~\cite{coq1}, \cite{kuz25}).\\
The magnetic excitation spectrum following from the GF (34)
consists of three branches - the acoustical spin wave, the optical
spin wave and the Stoner continuum~\cite{kuz25}. In the
hydrodynamic limit, $k \rightarrow 0$, $\omega \rightarrow 0$ the
GF (34) can be written as
\begin{equation}
<<S^{+}_{k} \vert S^{-}_{-k}>>^{0} =
\frac {2N^{-1/2} < \tilde S^{z}_{0}>}{
\omega - E(k)}
\end{equation}
where the acoustical spin wave energies are given by
\begin{equation}
E(k) = Dk^{2} = \frac {1/2 \sum_{q}(f_{q\uparrow} +
f_{q\downarrow})(\vec k \frac {\partial}{\partial \vec q})^{2}
\epsilon(\vec q) + (2\Delta)^{-1}\sum_{q}(f_{q\uparrow} -
f_{q\downarrow})(\vec k \frac {\partial}{\partial \vec q} \epsilon(\vec
q))^{2}}{
2N^{1/2}<S^{z}_{0}> + (n_{\uparrow} - n_{\downarrow})}
\end{equation}
and
\begin{equation}
< \tilde S^{z}_{0}> = <S^{z}_{0}> [ 1 +
\frac {(n_{\uparrow} - n_{\downarrow})}{
2N^{3/2} <S^{z}_{0}>}]^{-1}
\end{equation}
For  s.c. lattice the spin wave dispersion relation (36) becomes
\begin{eqnarray}
E(k) =
( 2N^{1/2}<S^{z}_{0}> +
(n_{\uparrow} - n_{\downarrow}))^{-1}  \\ \nonumber
( \frac {2t^{2}a^{2}}{\Delta}
\sum_{q} (f_{q\uparrow} - f_{q\downarrow})( k_{x} \sin (q_{x}a) +
k_{y} \sin (q_{y}a) + k_{z} \sin (q_{z}a))^{2} -
\nonumber \\
ta^{2} \sum_{q} (f_{q\uparrow} + f_{q\downarrow})(k^{2}_{x} \cos q_{x}a
+ k^{2}_{y} \cos q_{y}a
+ k^{2}_{z} \cos q_{z}a))\nonumber
\end{eqnarray}
In GMF approximation the density of itinerant electrons ( and the band
splitting $\Delta$) can be evaluated by solving the equation
\begin{equation}
n_{\sigma} = \sum_{k}<a^{+}_{k\sigma}a_{k\sigma}> = \sum_{k} [\exp
(\beta(\epsilon(k) + UN^{-1}n_{-\sigma} - JN^{-1/2}<S^{z}_{0}> -
\epsilon_{F})) + 1]^{-1}
\end{equation}
Hence, the stiffness constant $D$ can be expressed by the parameters of
the Hamiltonian (12).\\ The spectrum of the Stoner excitations is given
by~\cite{kuz25}
\begin{equation}
\omega_{k} = \epsilon(k+q) - \epsilon(q) + \Delta
\end{equation}
If we consider the optical spin wave branch then by direct calculation
one can easily show that
\begin{eqnarray}
E_{opt}(k) = E^{0}_{opt} + D(UE_{opt}/J\Delta - 1)k^{2} \nonumber \\
E^{0}_{opt} = J(n_{\uparrow} - n_{\downarrow}) + 2J<S^{z}_{0}>
\end{eqnarray}
From the equation (33) one also finds the GF of itinerant spin density
in the generalized mean field approximation
\begin{equation}
<<\sigma^{+}_{k}\vert \sigma^{-}_{-k}>>^{0}_{\omega}  =
\frac {\chi^{df}_{0}(k,\omega)}{1 - [ U - \frac {
2J^{2}<S^{z}_{0}>}{\omega - J( n_{\uparrow} - n_{\downarrow})}]
\chi^{df}_{0}(k,\omega)} \end{equation}
\subsection {Dyson Equation for d-f model}
The Dyson equation (10)  for the generalized $d-f$ model
has the following form
\begin{equation}  \hat G(k;\omega) =
\hat G_{0}(k;\omega) + \sum_{pq}
\hat G_{0}(p;\omega) \hat M(pq;\omega) \hat G(q;\omega) \end{equation}
The mass operator  $$\hat M(pq;\omega) = \hat
P^{(p)}(pq;\omega)$$ describes the inelastic (retarded) part of the
electron-phonon, electron-spin and electron-electron
interactions.  To obtain workable expressions for matrix elements of
the mass operator one should use the spectral theorem, inverse Fourier
transformation and make relevant approximation in the time correlation
functions.  The elements of the mass operator matrix $\hat M$ are
proportional to the higher-order GF of the following (conditional) form
$$ (^{(ir)}<<(S^{+})a_{k+p\sigma_{1}}a^{+}_{p+q\sigma_{2}}
a_{q\sigma_{2}}|(S^{-})a^{+}_{k+s\sigma_{3}}
a^{+}_{r\sigma_{4}} a_{r+s\sigma_{4}}>>^{(ir),p}) $$
For the explicit approximate calculation of the elements of the
mass operator it is convenient to write down the GFs in (44)
in terms of correlation functions by using the well-known spectral
theorem~\cite{tyab24}:
\begin{eqnarray}
(^{(ir)}<<(S^{+})a_{k+p\sigma_{1}}a^{+}_{p+q\sigma_{2}}a_{q\sigma_{2}}
|(S^{-})a^{+}_{k+s\sigma_{3}}
a^{+}_{r\sigma_{4}} a_{r+s\sigma_{4}}>>^{(ir),p}) =
\nonumber\\
{1 \over 2\pi}\int_{-\infty}^{+\infty}{d\omega' \over \omega - \omega'}
(\exp(\beta \omega') +1)     \int_{-\infty}^{+\infty}\exp(-i\omega't)dt
\nonumber\\
<(S^{-}(t))a^{+}_{k+s\sigma_{3}}(t)
a^{+}_{r\sigma_{4}}(t) a_{r+s\sigma_{4}}(t)|(S^{+})
a_{k+p\sigma_{1}}a^{+}_{p+q\sigma_{2}}a_{q\sigma_{2}}
>^{(ir),p})
\end{eqnarray}
Let us first consider the GF $<<A \vert \tilde A>>$ appearing in
$M_{11}$.  Further insight is gained if we select the suitable relevant
``trial" approximation for the correlation function on the r.h.s. of
(44). In this paper we show that the earlier formulations, based on the
decoupling or/and on diagrammatic methods can be arrived at from our
technique but in a self- consistent way. Clearly the choice of the
relevant trial approximation for the correlation function in (44) can be
done in a few ways.
The suitable or relevant approximations follow from the concrete
physical conditions of the problem under consideration. We consider
here for illustration the contributions from charge and spin
degrees of freedom by neglecting higher order contributions between the
magnetic excitations and charge density fluctuations as we did in
the theory of ferromagnetic ~\cite{kuz4} and antiferromagnetic
~\cite{kuz29},~\cite{kuz30} semiconductors. For example, a reasonable
and workable one may be the following approximation of two interacting
modes ~\cite{kuz3}
\begin{eqnarray}
<<A \vert \tilde A>>^{ir,p} \approx
\frac{J^{2}}{N^2 \pi^2} \sum_{kk_{1}k_{2}k_{3}k_{4}\sigma}
\int
\frac {d\omega_{1} d\omega_{2}}
{\omega - \omega_{1} - \omega_{2}} \\
F(\omega_{1},\omega_{2},) \nonumber \\
Im <<S^{+}_{k-k_{4}} \vert
S^{-}_{-k-k_{2}}>>_{\omega_{1}}
Im <<a^{+}_{k_{3} \sigma} a_{k_{3} + k_{4}\sigma} \vert
a^{+}_{k_{1} \sigma}a_{k_{1} + k_{2} \sigma} >>_{\omega_{2}}
\nonumber \\
F(\omega_{1},\omega_{2},) =
\frac { (\exp (\beta(\omega_{1} +
\omega_{2})) + 1}{(\exp (\beta \omega_{1}) - 1) (\exp (\beta
\omega_{2}) - 1)} \nonumber \end{eqnarray}
On the diagrammatic language this approximate expression results from
the neglecting of the vertex corrections.\\ The system of equations (43)
and (45) form a closed self-consistent system of equations.
In principle, one may use on the
r.h.s.  of (45) any workable first iteration-step forms of the GFs and
find a solution by repeated iterations.  It is most convenient to
choose as the first iteration step the following
approximations:
\begin{eqnarray}
Im <<S^{+}_{k-k_{4}} \vert
S^{-}_{-k-k_{2}}>>_{\omega_{1}} \approx
2\pi N^{-1/2}<S^{z}_{0}>\delta(\omega_{1} - E(k + k_{2})) \delta_{k_{4}
- k_{2}}; \\ Im <<a^{+}_{k_{3} \sigma} a_{k_{3} + k_{4}\sigma} \vert
a^{+}_{k_{1} \sigma}a_{k_{1} + k_{2} \sigma} >>_{\omega_{2}} \approx
\nonumber \\ \pi (f_{k_{3}\sigma} - f_{k_{1}\sigma}) \delta (\omega_{2}
+ \epsilon(k_{3}\sigma) - \epsilon(k_{3} + k_{4}\sigma))
\delta_{k_{3},k_{1} +k_{2}} \delta_{k_{1},k_{3} +k_{4}} \nonumber
\end{eqnarray} Then, using (46) in
(45), one can get an explicit expression for the $M_{11}$
\begin{equation}
<<A \vert \tilde A>>^{ir,p} \approx
\frac{2J^{2}}{N^2} \sum_{pq\sigma}
\frac {[1 + N(E(k+q)) - f_{p\sigma}]f_{p+q\sigma} +
N(E(k+q))f_{p\sigma}(1 - 2f_{p+q\sigma})} {\omega - E(k+q) -
\epsilon(p\sigma) + \epsilon(p+ q\sigma)}
\end{equation}
where
\begin{equation} \epsilon(k\sigma) =
\epsilon (k) + U<n_{-\sigma}>;  \quad
\nonumber
N(E(k)) = [\exp (\beta E(k)) - 1]^{-1}
\nonumber
\end{equation}
The calculations of the matrix elements $M_{12}$, $M_{21}$ and $M_{22}$
can be  done in the same manner, but with additional initial
approximation for phonon GF
\begin{equation}
<<Q_{k\nu} \vert Q^{+}_{k\nu}>> \approx (\omega^{2} -
\omega^{2}(k\nu))^{-1}
\end{equation}
It is transparent that the construction of the GF
$<<B_{p} \vert \tilde B_{q}>>$
will consist of contributions of the electron-phonon,
electron-magnon and electron-electron inelastic scattering.
$$<<B_{p} \vert \tilde B_{q}>> =
<<B_{p} \vert \tilde B_{q}>>^{ph-e} +
<<B_{p} \vert \tilde B_{q}>>^{m-e} +
<<B_{p} \vert \tilde B_{q}>>^{e-e}$$
As a result we
find the explicit expressions for the GFs in mass operator
\begin{eqnarray}
<<B_{p} \vert \tilde B_{q}>>^{ph-e} =
\frac{1}{2} \sum_{r\nu} \sum_{\alpha = \pm} \omega^{-1}(r\nu)
\nonumber \\ \Bigl ( \frac {[1 + N(\alpha \omega(r\nu)) -
f_{p+q+r\downarrow}]f_{p\uparrow} +
N(\alpha\omega(r\nu))f_{p+q+r\downarrow}(1 - 2f_{p\uparrow})}{\omega -
(\alpha \omega(r\nu) - \epsilon(p\uparrow) + \epsilon(p+
k+r \downarrow))} \nonumber \\ ((V^{\nu}(r,p+k))^{2} \delta_{q,p+k} -
V^{\nu}(r,p)V^{\nu}(r,p+k) \delta_{q,p+k+r}) +
\nonumber \\
\frac {[1 + N(\alpha \omega(r\nu)) -
f_{p+k\downarrow}]f_{p+r\uparrow} +
N(\alpha\omega(r\nu))f_{p+k\downarrow}(1 - 2f_{p+r\uparrow})}{\omega -
(\alpha \omega(r\nu) - \epsilon(p+r\uparrow) + \epsilon(p+
k \downarrow))} \nonumber \\ ((V^{\nu}(r,p))^{2} \delta_{q,p+k} -
V^{\nu}(r,p)V^{\nu}(r,p+k) \delta_{q,p+k+r}) \Bigr )
\end{eqnarray}
The contribution from inelastic electron-magnon scattering is
given by
\begin{eqnarray}
<<B_{p} \vert \tilde B_{q}>>^{m-e} = -
\frac{2J^{2}}{N^{2}}<S^{z}_0> \sum_{r}
\nonumber \\ \Bigl ( \frac {[1 + N(
E(r)) - f_{p+k+r\uparrow}]f_{p\uparrow} +
N(E(r))f_{p+k+r\uparrow}(1 - 2f_{p\uparrow})}{\omega -
(E(r) - \epsilon(p\uparrow) + \epsilon(p+
k+r \uparrow))} \nonumber \\
 +
\nonumber \\
\frac {[1 + N(E(r)) -
f_{p+k\downarrow}]f_{p+r\downarrow} +
N(E(r))f_{p+k\downarrow}(1 - 2f_{p+k\downarrow})}{\omega -
(E(r) - \epsilon(p+r\downarrow) + \epsilon(p+
k \downarrow))}
\Bigr ) \delta_{q,p+k}
\end{eqnarray}
The term due to the electron-electron inelastic scattering processes
becomes
\begin{eqnarray}
<<B_{p} \vert \tilde B_{q}>>^{e-e} =
\frac{U^{2}}{N^{2}}
\nonumber \\ \Bigl ( \sum_{rs}
[ \frac {(1
- f_{p+k\downarrow})(1 -
f_{r+s\downarrow})f_{r\downarrow}f_{p+k\uparrow} +
f_{p+k\downarrow}f_{r+s\downarrow}(1 -
f_{r\downarrow})(1 - f_{p+s\uparrow})}{\omega -
(\epsilon(p+k\downarrow) - \epsilon(p+s\uparrow) + \epsilon(r+s
\downarrow) - \epsilon(r\downarrow))}  + \nonumber \\
 \frac {(1
- f_{p+k+s\downarrow})(1 -
f_{r-s\uparrow})f_{r\uparrow}f_{p\uparrow} +
f_{p+k+s\downarrow}f_{r-s\uparrow}(1 -
f_{r\uparrow})(1 - f_{p\uparrow})}{\omega -
(\epsilon(p+k+n\downarrow) - \epsilon(p\uparrow) + \epsilon(r-s
\uparrow) - \epsilon(r\uparrow))} ]  - \nonumber \\
\sum_{r} [ \frac {(1
- f_{q\downarrow})(1 -
f_{p+k\downarrow})f_{r\downarrow}f_{p+q-r\uparrow} +
f_{q\downarrow}f_{p+k\downarrow}(1 -
f_{r\downarrow})(1 - f_{p+q-m\uparrow})}{\omega -
(\epsilon(p+k\downarrow) + \epsilon(q\downarrow) -\epsilon(r\downarrow)
- \epsilon(p+q-r\uparrow))}]  + \nonumber \\
\frac {(1
- f_{q+r\downarrow})(1 -
f_{p-r\uparrow})f_{p\uparrow}f_{q-k\uparrow} +
f_{q+r\downarrow}f_{p-r\uparrow}(1 -
f_{p\uparrow})(1 - f_{q-k\uparrow})}{\omega -
(\epsilon(q+r\downarrow) + \epsilon(p-r\uparrow)
-\epsilon(p\uparrow) - \epsilon(q-k\uparrow))} ]
\Bigr ) \delta_{q,p+k} \end{eqnarray}
In the same way for off-diagonal contributions we find
\begin{eqnarray}
<<A \vert \tilde B_{q}>> = -
\frac{2J^{2}}{N^{2}}<S^{z}_0> \sum_{r}
\nonumber \\ \Bigl ( \frac {[1 + N(
E(r)) - f_{q+r\uparrow}]f_{q-k\uparrow} +
N(E(r))f_{q+r\uparrow}(1 - 2f_{q-k\uparrow})}{\omega -
(E(r) + \epsilon(q+r\uparrow) - \epsilon(q
-k \uparrow))} \nonumber \\
 +
\nonumber \\
\frac {[1 + N(E(r)) -
f_{q\downarrow}]f_{q+r-k\downarrow} +
N(E(r))f_{q\downarrow}(1 - 2f_{q+r-k\downarrow})}{\omega -
(E(r) - \epsilon(q+r-k\downarrow) + \epsilon(q
\downarrow))}
\Bigr )
\end{eqnarray}
And we have $<<B_{p} \vert \tilde A>> = <<A \vert \tilde B_{p+k}>>$.
\subsection{Self-Energy and Damping}
Finally we turn to the calculation of the damping.
To find the damping of the quasiparticle states in the general case,
one needs to find the matrix elements of the mass-operator in (43).
Thus we have \begin{equation} \pmatrix{ \hat G_{11}&\hat G_{12}\cr \hat
G_{21}&\hat G_{22}\cr} =\left[ \pmatrix{ \hat G_{011}&\hat G_{012}\cr
\hat G_{021}&\hat G_{022}\cr}^{-1} - \pmatrix{\hat M_{11}&\hat M_{12}\cr
\hat M_{21}&\hat M_{22}\cr}\right]^{-1} \end{equation}
From this matrix equation we have\\
\begin{eqnarray}
M_{11} = \frac{J^2}{N\Omega^2_2}<<A \vert \tilde A>>
; \nonumber \\  M_{21} = \frac{J}{\Omega_2 N^{3/2} \chi^{df}_0}
\sum_{p} (\omega_{p,k})^{-1} <<B_{p} \vert \tilde A>>; \nonumber  \\
M_{12} = \frac{J}{\Omega_2 N^{3/2} \chi^{df}_{0}} \sum_{q}
(\omega_{q,p})^{-1}<<A \vert \tilde B_{q}>>; \nonumber \\  M_{22} =
\frac{ 1}{N^2 (\chi^{df}_{0})^2} \sum_{pq} (\omega_{p,k}
\omega_{q,k})^{-1} <<B_{p} \vert \tilde B_{q}>>; \end{eqnarray}
With (54) and (55) the GF $\hat G$ becomes
\begin{equation}
\hat G = \frac{1}{det(\hat G^{-1}_{0} - \hat M)}
\pmatrix{
-\frac{(1 - U\chi^{df}_{0} )}{N\chi^{df}_{0}} - M_{22}
& -(\frac {J}{N^{1/2}} - M_{12})\cr
-(\frac {J}{N^{1/2}} - M_{21})
&(\frac {J}{N^{1/2}} \frac{\Omega_{1}}{\Omega_{2}} - M_{11})\cr}
\end{equation}
Let us estimate the damping of magnetic excitations. From (56) we find
\begin{equation}
<<S^{+}_{k} \vert S^{-}_{-k}>>_{\omega} = \frac {1}{(G^{11}_{0})^{-1} -
\Sigma(k,\omega)}
\end{equation}
Here the self-energy $\Sigma$ is given by
\begin{eqnarray}
\Sigma(k,\omega) = M_{11} + \frac {J^{2} \chi^{df}_{0}}{1 -
U\chi^{df}_{0}} - \nonumber \\ (JN^{-1/2} - M_{12})(JN^{-1/2} - M_{21})
N\chi^{df}_{0} \Bigl ((1 - U\chi^{df}_{0}) + M_{22}N\chi^{df}_{0} \Bigr
)^{-1}
\end{eqnarray}
Let us consider the damping of the acoustical magnons. Considering only
the linear terms in the matrix elements of the mass operator in (58),
we find for small $\vec k$ and $\omega$ \begin{equation} <<S^{+}_{k}
\vert S^{-}_{-k}>>_{\omega} \approx \frac {2N^{-1/2}<\tilde
S^{z}_{0}>}{\omega - E(k) - 2N^{-1/2}<\tilde
S^{z}_{0}>\Sigma(k,\omega)} \end{equation} where \begin{eqnarray}
\Sigma(k,\omega) \approx M_{11} + (M_{12} + M_{21}) \frac {JN^{1/2}
\chi^{df}_{0}}{1 - U\chi^{df}_{0}} +  \frac {J^{2}N
(\chi^{df}_{0})^{2}}{(1 -
U\chi^{df}_{0})^{2}} M_{22} \end{eqnarray}
Then the spectral density of the spin-wave excitations will be given as
\begin{eqnarray}
 -{1 \over \pi}Im
G^{11}(k, \omega + i\varepsilon) = -{1 \over \pi}Im
<<S^{+}_{k} \vert S^{-}_{-k}>>_{\omega} = \nonumber \\ \frac
{2N^{-1/2}<\tilde S^{z}_{0}> \Gamma (k,\omega)}{(\omega - E(k) -
\Delta(k,\omega))^{2} + \Gamma^{2}(k,\omega)}
\end{eqnarray}
where
\begin{eqnarray}
\Delta(k, \omega) =
2N^{-1/2}<\tilde S^{z}_{0}> Re \Sigma (k,\omega) \nonumber \\
\Gamma (k, \omega) = 2N^{-1/2}<\tilde S^{z}_{0}> Im \Sigma (k,\omega +
\varepsilon)
\end{eqnarray}
describes the shift and the damping of the magnons, respectively.\\
Finally we estimate the temperature dependence of $\Gamma(k,\omega)$
due to the mass operator terms in (58). Considering the first
contribution in (58) we get
\begin{eqnarray}
Im M_{11} = Im
<<A \vert \tilde A>>_{\omega} \approx
J^{2}< S^{z}_{0}> \sum_{pq\sigma}  \Bigl ( (1 +
N(E(k+q)) - f_{p\sigma})f_{p+q\sigma} + \nonumber \\
N(E(k+q))f_{p\sigma}(1 - 2f_{p+q\sigma}) \Bigr )
\delta(\omega - E(k+p) + \epsilon(p+q) - \epsilon(p)) \end{eqnarray}
Using the standard relations
\begin{eqnarray}
\sum_{pq} \rightarrow {V^{2} \over (2\pi)^{6}} \int d^{3}p \int d^{3}q
\nonumber \\
N(E(q)) \vert _{q \rightarrow 0} = (\exp (\beta Dq^{2}) - 1 )^{-1}
\end{eqnarray}
we find
\begin{eqnarray}
Im M_{11} \sim
J^{2}< S^{z}_{0}>
{V^{2} \over (2\pi)^{6}} 2\pi \int d^{3}p \int^{q_{max}}_{0} q^{2} dq
\int d(\cos \Theta) \nonumber \\ \tilde F(f_{p\sigma}, N(E(k+p)) \frac
{\delta (\cos \Theta - \cos \Theta_{0})}{\vert {\partial \epsilon \over
\partial p} \vert q} \nonumber \\ \sim {1 \over 2\beta D} \int ^{\beta
E_{max}}_{0} dx \frac {1}{\exp x - 1} \sim T \end{eqnarray}
The other contributions to $\Gamma(k,\omega)$ can be treated in the
same way, where $M_{12}$, $M_{21}$ and electron-magnon contribution to
$M_{22}$ are proportional to $T$, too. For the electron-phonon
contribution to $M_{22}$ we find
\begin{eqnarray}
Im M^{ph}_{22} = Im
<<B_{p} \vert \tilde B_{q}>>^{ph}_{\omega}
\sim {1 \over \beta^{3} } \int
x^{2} dx \frac {1}{\exp x - 1} \sim T^{3} \end{eqnarray}
Hence, the damping of the acoustical magnons at low temperatures can be
written as
\begin{equation}
\Gamma(k,\omega) \vert _{k,\omega \rightarrow 0} \sim \Gamma_{1} +
\Gamma_{2} T + \Gamma_{3} T^{3}
\end{equation}
where the coefficients $\Gamma_{i}$, $( i = 1,2,3)$ vanish for
$k = \omega = 0$, and furthermore for $J = 0$.\\
\subsection{Charge dynamics of d-f model}
To describe the quasiparticle charge dynamics or dynamics of carriers
of the $d-f$ model (12) we should consider the equation of motion for
the  GF of the form
\begin{equation}
G_{k\sigma} = <<a_{k\sigma} \vert a^{+}_{k\sigma}>>
\end{equation}
Performing the first time differentiation of (68) we find
\begin{eqnarray}
(\omega - \epsilon(k))G_{k\sigma} = 1 + {U \over N} \sum_{pq}
<<a^{+}_{p+q-\sigma}a_{p-\sigma}a_{k+q\sigma} \vert a^{+}_{k\sigma}>> -
\nonumber \\
\frac {J}{N^{1/2}} \sum_{q} \Bigl ( <<S^{-\sigma}_{-q} a_{k+q-\sigma}
\vert a^{+}_{k\sigma}>> + z_{\sigma}<<S^{z}_{-q}a_{k+q\sigma} \vert
a^{+}_{k\sigma}>> \Bigr) + \nonumber \\
\sum_{q\nu\alpha} V^{\alpha}(k-q,k) <<a_{k-q\sigma} Q_{q\nu} \vert
a^{+}_{k\sigma}>> \end{eqnarray}
Following the previous consideration we should introduce the
irreducible GFs and perform the differentiation of the higher-order GFs
on second time. Using this approach the the equation of motion (69) can
be exactly transformed into the Dyson equation
\begin{equation}   G_{k\sigma}(\omega) =
G^{0}_{k\sigma}(\omega) +
G^{0}_{k\sigma}(\omega) M_{k\sigma}(\omega) G_{k\sigma}(\omega)
\end{equation}
where
\begin{eqnarray}
G^{0}_{k\sigma} = (\omega - \epsilon^{0}(k\sigma))^{-1} \nonumber \\
\epsilon^{0}(k\sigma) = \epsilon(k) -z_{\sigma} {1 \over
N^{1/2}} <S^{z}_{0}> + {U \over N} n_{-\sigma}
\end{eqnarray}
Here the mass operator has the following exact representation
\begin{equation}
M_{k\sigma}(\omega) = M_{k\sigma}^{ee}(\omega) +
M_{k\sigma}^{e-m}(\omega) + M_{k\sigma}^{e-ph}(\omega)
\end{equation}
where
\begin{eqnarray}
M_{k\sigma}^{ee}(\omega) = {U^2 \over N^2} \sum_{pqrs}
(^{(ir)}<<a^{+}_{p+q-\sigma}a_{p-\sigma}
a_{p+q\sigma}|a^{+}_{r+s-\sigma}
a_{r-\sigma} a^{+}_{k-s\sigma}>>^{(ir),p}) \\
M_{k\sigma}^{e-m}(\omega) = {J^2 \over N} \sum_{qs} \Bigl (
(^{(ir)}<<S^{-\sigma}_{-q}
a_{k+q-\sigma}|S^{\sigma}_{s}a^{+}_{k+s-\sigma}>>^{(ir),p}) + \nonumber
\\ (^{(ir)}<<S^{z}_{-q}
a_{k+q\sigma}|S^{z}_{s}a^{+}_{k+s\sigma}>>^{(ir),p}) \Bigl ) \\
M_{k\sigma}^{e-ph}(\omega) = \sum_{q\nu\alpha} \sum_{s\mu\alpha'}
V^{\alpha}_{q\nu}(p-q,p)V^{\alpha'}_{s\mu}(p,p+q)
(^{(ir)}<<Q_{q\nu}
a_{p-q\sigma}|Q_{s\mu}a^{+}_{p+q\sigma}>>^{(ir),p})
\end{eqnarray}
As previously, we express the GF in terms of the correlation functions.
In order to calculate the mass operator self-consistently we shall use
the "pair" approximation~\cite{kuz15},\cite{kuz20} for the $M^{ee}$ and
approximation of two interacting modes for $M^{e-m}$ and $M^{e-ph}$
\cite{kuz4}, \cite{kuz26}. Then the corresponding expressions can be
written as
\begin{eqnarray}
M^{ee}_{k\sigma}(\omega) =
\frac{U^{2}}{N^2} \sum_{pq}
\int
\frac {d\omega_{1} d\omega_{2} d\omega_{3}}
{\omega + \omega_{1} - \omega_{2} - \omega_{3}} \nonumber \\
F^{ee}(\omega_{1},\omega_{2}, \omega_{3}) \nonumber \\
g_{p+q,-\sigma}(\omega_{1})g_{k+p,\sigma}(\omega_{2})g_{p,-\sigma}(\omega_{3})
\end{eqnarray}
where
$$ g_{k\sigma}(\omega) = {-1 \over \pi}Im <<a_{k\sigma} \vert
a^{+}_{k\sigma}>>_{\omega + \varepsilon}$$ \\ and $$
F^{ee}(\omega_{1},\omega_{2}, \omega_{3}) = (f(\omega_{1})(1 -
f(\omega_{2}) -f(\omega_{3})) + f(\omega_{2})f( \omega_{3}))$$
Let us consider now the spin-electron inelastic scattering. As
previously, we shall neglect of the vertex corrections, i.e.
correlation between the propagations of the charge and spin
excitations. Then we obtain from (74)
\begin{eqnarray}
M^{e-m}_{k\sigma}(\omega) =
\frac{J^{2}}{N} \sum_{q}
\int
\frac {d\omega_{1} d\omega_{2}}
{\omega - \omega_{1} - \omega_{2}}
F^{em}(\omega_{1},\omega_{2})  \nonumber \\
\Bigl ( g_{k+p,-\sigma}(\omega_{2})({-1 \over \pi}Im
<<S^{\sigma}_{-q} \vert
S^{-\sigma}_{q}>>_{\omega_{1}}) + g_{k+p,\sigma}(\omega_{2})
({-1 \over \pi}Im
<<S^{z}_{q} \vert
S^{z}_{-q}>>_{\omega_{1}}) \Bigr )
\end{eqnarray}
where
$$F^{em}(\omega_{1},\omega_{2}) = (1 + N(\omega_{1}) -
f(\omega_{2}))$$ And finally we shall find the similar expression
for electron-phonon inelastic scattering contribution (75)
\begin{eqnarray}
M^{e-ph}_{k\sigma}(\omega) =
\sum_{q\nu} \vert V_{\nu}(\vec p - \vec q, \vec p) \vert ^{2}
\int
\frac {d\omega_{1} d\omega_{2}}
{\omega - \omega_{1} - \omega_{2}}
F^{e-ph}(\omega_{1},\omega_{2})  \nonumber \\
g_{p-q,\sigma}(\omega_{1})({-1 \over \pi}Im
<<Q_{q\nu} \vert
Q^{+}_{q\nu}>>_{\omega_{2}})
\end{eqnarray}
where
$$F^{e-ph}(\omega_{1},\omega_{2}) = (1 + N(\omega_{2}) -
f(\omega_{1}))$$
Equations (70), (76), (77) and (78) form a closed self-consistent
system of equations for one-fermion GF of  the carriers for a
generalized spin-fermion model. To find explicit expressions for the
mass operator (72) we choose for the first iteration step in (76) - (78)
the following trial approximation
\begin{equation}
g_{k\sigma}(\omega) = \delta(\omega - \epsilon^{0}(k\sigma))
\end{equation}
Then we find
\begin{eqnarray}
M^{ee}_{k\sigma}(\omega) =
\frac{U^{2}}{N^2} \sum_{pq}
\frac{
f_{p+q\sigma}(1 -
f_{k+p\sigma} -f_{q-\sigma}) + f_{k+p\sigma}f_{q-\sigma}}
{\omega + \epsilon^{0}(q,-\sigma) - \epsilon^{0}(p+q,\sigma) -
\epsilon^{0}(k+p\sigma)}
\end{eqnarray}
For the initial trial approximation for the spin GF we take the
expression (46) in the following form
\begin{eqnarray}
{-1 \over \pi} Im <<S^{\sigma}_{q} \vert
S^{-\sigma}_{-q}>> \approx z_{\sigma}
(2 N^{-1/2}<S^{z}_{0}>) \delta(\omega - z_{\sigma} E(q))
\end{eqnarray}
Then we obtain~\cite{kuz4}
\begin{eqnarray}
M^{e-m}_{k\uparrow}(\omega) =
\frac{2J^{2}<S^{z}_{0}>}{N^{3/2}} \sum_{q}
\frac {f_{k+q,\downarrow} + N(E(q))}
{\omega - \epsilon^{0}(k+q,\downarrow) - E(q)}~; \nonumber \\
M^{e-m}_{k\downarrow}(\omega) =
\frac{2J^{2}<S^{z}_{0}>}{N^{3/2}} \sum_{q}
\frac {1- f_{k-q,\uparrow} + N(E(q))}
{\omega - \epsilon^{0}(k-q,\uparrow) - E(q)}
\end{eqnarray}
This result is written for the low temperature region, when one
can drop the contributions from the dynamics of longitudinal
$(zz)$ GF
which is essential at high temperatures and in some special cases.\\
In order to calculate the electron-phonon term (78) we need to take as
initial approximation the expressions (49) and (79). We then get
\begin{eqnarray}
M^{e-ph}_{k\sigma}(\omega) =
\sum_{q\nu} \frac {\vert V_{\nu}(\vec p - \vec q, \vec p) \vert ^{2}}
{2\omega(q\nu)} \Bigl (
\frac {1- f_{k-q,\sigma} + N(\omega(q\nu))}
{\omega - \epsilon^{0}(k-q,\uparrow) - \omega(q\nu)} +
\frac { f_{k-q,\sigma} + N(\omega(q\nu))}
{\omega - \epsilon^{0}(k-q,\uparrow) + \omega(q\nu)} \Bigr )
\end{eqnarray}
where
\begin{equation}
\vert V_{\nu}(\vec p - \vec q, \vec p) \vert ^{2} =
\sum_{\alpha} \frac {4q^{0}t^{2}}{NM}(\sin \vec a_{\alpha} \vec p -
\sin \vec a_{\alpha}(\vec p - \vec q) )^{2} \vert e^{\alpha}_{\nu}(\vec
q ) \vert^{2}
\end{equation}
Then analysis of the electron-phonon term can be done as in
Ref.~\cite{kuz26}.\\
For the fully self-consistent solution of the problem the phonon
GF can be easily calculated too. The final result is
\begin{equation}
<<Q_{k\nu} \vert Q^{+}_{k\nu}>> = (\omega^{2} -
\omega^{2}(k\nu) - \Pi_{k\nu}(\omega))^{-1}
\end{equation}
where the polarization operator $\Pi$ has the form
\begin{eqnarray}
\Pi_{k\nu}(\omega) =
\sum_{q\sigma} \vert V_{\nu}(\vec q - \vec k, \vec q) \vert
^{2}  \frac { f_{q-k,\sigma} - f_{q\sigma}}
{\omega + \epsilon^{0}(q-k,\sigma) - \epsilon^{0}(q\sigma)}
\end{eqnarray}
The above expressions were derived in the self-consistent way for the
generalized spin-fermion model and for finite temperatures. \\It is
important to note that to investigate the spin and charge dynamics in
doped manganite perovskites the scheme described above should be
modified to take into account the strong Hund rule coupling in these
systems but it deserve of separate consideration.  In the present paper
to show clearly the advantage of the IGF approach we shall consider
another interesting example, the dynamics of carriers for the
Kondo-Heisenberg model.

\section {Dynamics of Carriers in the Spin-Fermion Model.  }
\subsection{Hole Dynamics in Cuprates}
To show the specific behaviour of the carriers in the framework of
spin-fermion model  we shall consider a dynamics of holes in HTSC
cuprates. A vast amount of theoretical searches for the relevant
mechanism of high temperature superconductivity  deals with the
strongly correlated electron models~\cite{kuz10}.  Much
attention has been devoted to the formulation of successful theory of
the electrons (or holes) propagation in the $CuO_{2}$ planes in copper
oxides. In particular, much efforts have been done to describe
self-consistently the hole propagation in the doped 2D quantum
antiferromagnet~\cite{kane31} - \cite{feng44}.  The understanding of
the true nature of the electronic states in HTSC are one of the central
topics of the current experimental and theoretical efforts in the
field~\cite{kuz10},\cite{su37}.
Theoretical description  of
strongly correlated fermions on two-dimensional lattices and the hole
propagation in the antiferromagnetic background still remains
controversial. The role of quantum spin
fluctuations was found to be quite crucial for the hole propagation.
The essence of the problem  is in the inherent interaction (and
coexistence) between charge and spin degrees of freedom which are
coupled in a self-consistent way. The propagating hole perturbs the
antiferromagnetic background and move then together with the distorted
underlying region. There were many
attempts to describe adequately  this motion.
However, a definite proof of the fully adequate mechanism for the
coherent  propagation of the hole is still lacking. In this paper we will
analyse the physics of the doped systems and the true nature of
carriers in the 2D antiferromagnetic background from the many-body
theory point of view. The dynamics of the charge degrees of freedom
for the $CuO_{2}$ planes  in copper oxides will be described in the
framework of the spin-fermion (Kondo-Heisenberg)
model~\cite{pre7},~\cite{pre35}, using the approach described in
Section 3. \subsection{Hubbard model and t-J model} Before
investigating the Kondo-Heisenberg model it is instructive to consider
the $t-J$ model very briefly. The $t-J$ model is a special development
of the spin-fermion model approach which reflect the specific of
strongly correlated systems. To remind this let us consider first the
Hubbard model~\cite{hub28}.  \\The model Hamiltonian which is usually
refered as to Hubbard Hamiltonian is given by \begin{equation} H =
\sum_{ij\sigma}t_{ij}a^{+}_{i\sigma}a_{j\sigma} + \frac {U}{2}
\sum_{i\sigma}n_{i\sigma}n_{i-\sigma} \end{equation} For the strong
coupling limit, when Coulomb integral U $\gg$ W, where W is the
effective bandwidth, the Hubbard Hamiltonian is reduced in the
low-energy sector to t-J model Hamiltonian of the form \begin{equation}
H = \sum_{ij\sigma}(t_{ij}(1 - n_{i-\sigma})a^{+}_{i\sigma}
a_{j\sigma}(1 - n_{j-\sigma}) + H.C.) +
J\sum_{ij}S_{i}S_{j}
\end{equation}
This Hamiltonian play an important role in the theory of HTSC.
Let us consider the carrier motion. The hopping at
half-filling is impossible and this model describe the planar
Heisenberg antiferromagnet. The most interesting problem is the behaviour
of this system when the doped holes are added. In the $t-J$ model
($U \rightarrow \infty$) doped holes can move only in the projected
space, without producing doubly occupied configurations
($<n_{\uparrow}> +
<n_{\downarrow}> \leq 1$). There is then a strong competition between the
kinetic energy of the doped carriers and the magnetic order present
in the system.
According to Ref.~\cite{hu34}, it is
possible to rewrite first term in (88) in the following form
\begin{equation}
H_{t} = t\sum_{<ij>}(a^{+}_{i\uparrow}S^{-}_{i}S^{+}_{j}a_{j\uparrow}
+ a^{+}_{i\downarrow}S^{+}_{i}S^{-}_{j}a_{j\downarrow} + h.c.)
\end{equation}
This form show clearly the nature hole-spin correlated motion over
antiferromagnetic background. It is follows from (89) that
to describe in a self-consistent way a correlated motion of a carrier
one need to consider the following  matrix GF Function:
\begin{equation}
G(i,j) = \pmatrix{
<<a_{i\uparrow}|a^{+}_{i\uparrow}>>&<<a_{i\uparrow}|a^{+}_{j\downarrow}>>&
<<a_{i\uparrow}|S^{+}_{j}>>&<<a_{i\uparrow}|S^{-}_{j}>>\cr
<<a_{i\downarrow}|a^{+}_{j\uparrow}>>&<<a_{i\downarrow}|
a^{+}_{j\downarrow}>>&<<a_{i\downarrow}|S^{+}_{j}>>&<<a_{i\downarrow}|
S^{-}_{j}>>\cr
<<S^{-}_{i}|a^{+}_{j\uparrow}>>&<<S^{-}_{i}|a^{+}_{j\downarrow}>>&
<<S^{-}_{i}|S^{+}_{j}>>&<<S^{-}_{i}|S^{-}_{j}>>\cr
<<S^{+}_{i}|a^{+}_{j\uparrow}>>&<<S^{+}_{i}|a^{+}_{j\downarrow}>>&
<<S^{+}_{i}|S^{+}_{j}>>&<<S^{+}_{i}|S^{-}_{j}>>\cr}
\end{equation}
It may be shown after  straightforward but tedious manipulations
by using IGF method of Section 2 that the equation of motion (2)
for the GF (90) can be rewritten as a Dyson equation (10)
\begin{equation} G(i,j;\omega) =
G_{0}(i,j;\omega) + \sum_{mn}G_{0}(i,m;\omega)M(m,n;\omega)
G(n,j;\omega)
\end{equation}
The algebraic structure of the full GF in (91) which follows from (11)
is rather complicated.
For clarity, we illustrate some features by means of simplified problem.
\subsection{Hole Spectrum of $t-J$ model}
It is well known~\cite{su37},\cite{feng44} how to write
down the special ansatz for fermionic operator as a composite
operator of dressed hole operator and spin operator
for the case $J \gg t$. The hole operator $h_{i}$
corresponding to fermion operator $a^{+}_{i\sigma}$ on the spin-up
sublattice using the ansatz $a^{+}_{i\uparrow} = h_{i}S^{-}_{i}$ and
similarly for spin-down sublattice have been introduced ( for a recent
discussion see e.g. Ref.~\cite{feng44}).  Then the Hamiltonian (89)
obtain the form \begin{equation} H_{t} = t\sum_{ij}I_{ij}
h^{+}_{j}h_{i}(b^{+}_{i} + b_{j}) \end{equation} Here $b_{i}$ and
$b^{+}_{j}$ are the boson operators, which results from the
Holstein-Primakoff transformation of spins into bosons.  Equation (92)
is not convenient form because of its non-diagonal structure.  Caution
should be exercised because the new vacuum is a distorted Neel
vacuum.\\ The equation of motion (2) and (3) for the hole GF can be
written in the following form
\begin{eqnarray} \omega
<<h_{j}|h^{+}_{k}>> - t\sum_{n}I_{jn}<B_{nj}><<h_{n}|h^{+}_{k}>> =
\nonumber \\
\delta_{jk} + t\sum_{n}I_{jn}(^{ir}~<<h_{n}B_{nj}|h^{+}_{k}>>)
\end{eqnarray}
Here $B_{nj} = (b^{+}_{n} + b_{j})$. The "mean-field" GF (6) is
defined by
\begin{equation}
\sum_{i}(\omega \delta_{ij} - tI_{ij}<B_{ji}>)G_{0}(i,k;\omega) = \delta_{jk}
\end{equation}
Note, that "spin distortion" $<B_{mn}>$ does not depend on $(R_{m} - R_{n})$.
Then the Dyson equation (91) becomes
\begin{equation}
G(g,k) = G_{0}(g,k) +\sum_{jl}G_{0}(g,j)M(j,l)G(l,k)
\end{equation}
where self-energy operator is given by
\begin{equation}
M(j,l) = t^{2}\sum_{mn}I_{jn}
(^{ir}<<h_{n}B_{nj}|h^{+}_{m}B_{lm}>>^{ir}) I_{ml} \end{equation} The
standard IGF-method's prescriptions for the approximate calculation of
the self-energy ( c.f. Section 3.4) , can be written in the form
\begin{eqnarray} M(j,l;\omega) =
t^{2}\sum_{mn}I_{jn}I_{ml}\int_{-\infty}^{+\infty} d\omega_{1}
d\omega_{2} \frac {1 + N(\omega_{1}) - f(\omega_{2})}{\omega -
\omega_{1} - \omega_{2}}\\ \nonumber (\frac {1}{\pi}
Im<<B_{nj}|B_{lm}>>_{\omega_{1}})(\frac {1}{\pi} ImG(n,m;\omega_{2}))
\end{eqnarray}
It is worthy to note that the "mass" operator (97) is proportional
to $t^{2}$ . The standard iterative self-consistent procedure of
IGF approach for the calculation of mass operator encounter the
need of choosing as a first iteration "trial" solution the
non-diagonal initial spectral function $ImG_{0}$; in another
words, there are no reasonable "zero-order" approximation for
dynamical behaviour. The initial hole GF in paper~\cite{su33} was
defined as \begin{equation} G_{0}(j,k;\omega) =
\frac{\delta_{jk}}{\omega + i\epsilon} \end{equation} which
corresponds to static hole, without dispersion.  In contrast, the
approximation for the magnon GF yield the momentum distribution of
a free magnon gas. After integration in (97) , the mass operator
is given by an expression quite similar to the one encountered in
papers~\cite{su32}, where the Bogoliubov-de Gennes equations has
been derived.  It can be checked that the present set of equations
(95) - (97) gives the finite temperature generalisation of the
results \cite{su33}. As we just mentioned, one of its main merits
is that it enables one to see clearly the "composite" nature of
the hole states in an antiferromagnetic background, but in the
quasi-static limit. The recent analysis~\cite{dot42},\cite{feng44}
show that the difficulties of the consistent description of the
coherent hole motion within $t-J$ model are rather intrinsic
properties of the model and of the very complicated many-body
effects. From this point of view it will be instructive to
reanalyse the less complicated model Hamiltonian, in spite of the
fact that its applicability has been determined as the less
reliable.\subsection{Kondo-Heisenberg Model} As far as the
$CuO_{2}$-planes in the copper oxides are concerned, it was
argued~\cite{pre7},\cite{pre35} that a relatively reasonable
workable model  with which one can discuss the dynamical
properties of charge and spin subsystems is the spin-fermion ( or
Kondo-Heisenberg) model~\cite{pre7}. This model allows for motion
of doped holes and results from d-p model Hamiltonian.  We
consider the interacting hole-spin model for a copper-oxide planar
system described by the Hamiltonian \begin{equation} H = H_{t} +
H_{K} + H_{J}
\end{equation} where $H_{t}$ is the doped hole Hamiltonian
\begin{equation} H_{t} = - \sum_{<ij>
\sigma}(ta^{+}_{i\sigma}a_{j\sigma} + H.C.) =
\sum_{k\sigma}\epsilon(k)a^{+}_{k\sigma}a_{k\sigma}
\end{equation}
where $a^{+}_{i\sigma}$ and $a_{i\sigma}$ are the creation and
annihilation second quantized fermion operators, respectively for
itinerant carriers with energy spectrum
\begin{equation}
\epsilon (q) = -4t cos(1/2q_{x})cos(1/2q_{y}) = t\gamma_{1}(q).
\end{equation}
The term $H_{J}$ in (99) denotes Heisenberg superexchange Hamiltonian
\begin{equation}
H_{J} = \sum_{<mn>}J{\vec S_{m}}{\vec S_{n}} = \frac{1}{2N} \sum_{q}
J(q){\vec S_{q}}{\vec S_{-q}}
\end{equation}
Here ${\vec S_{n}}$ is the operator for a spin at copper site ${\vec
r_{n}}$ and J is the n.n. superexchange interaction
\begin{equation}
J(q) = 2J[cos(q_{x}) + cos(q_{y})] = J\gamma_{2}(q)
\end{equation}
Finally, the hole-spin (Kondo type) interaction $H_{K}$ may be written
as (for one doped hole)
\begin{equation}
H_{K} = \sum_{i}K{\vec \sigma_{i}}{\vec S_{i}} =
N^{-1/2}\sum_{kq}\sum_{\sigma}K(q)[S^{-\sigma}_{-q}a^{+}_{k\sigma}
a_{k+q-\sigma} + z_{\sigma}S^{z}_{-q}a^{+}_{k\sigma}a_{k+q\sigma}]
\end{equation}
This part of the Hamiltonian was written as the sum of a
dynamic(or spin-flip) part and a static one. Here K is hole-spin
interaction energy
\begin{equation}
K(q) = K[cos(1/2 q_{x}) + cos(1/2 q_{y})] = K\gamma_{3}(q)
\end{equation}
We start in this
paper with the one doped hole model (99), which is considered to have
captured the essential physics of the multi-band strongly correlated
Hubbard model in the most interesting parameters regime $t > J, \vert K
\vert$.  We apply the IGF method to this 2D variant of the spin-fermion
model.  It will be shown that we are able to give a much more
detailed and self-consistent description of the fermion and spin
excitation spectra than in papers~\cite{fur8} - \cite{ino9}, including
the damping effects and finite lifetimes.\\ For a recent discussion of
the one-dimensional Kondo-Heisenberg model and the classification of
the ground-state phases of this model in the context of a fixed-point
strategy see Ref.~\cite{emer45}. \subsection{Hole Dynamics in the
Kondo-Heisenberg Model} The two-time thermodynamic GFs to
be studied here are given by \begin{equation} G(k\sigma, t - t') =
<<a_{k\sigma}(t),a^{+}_{k\sigma}(t')>> = -i\theta (t -
t')<[a_{k\sigma}(t),a^{+}_{k\sigma}(t')]_{+}> \end{equation}
\begin{equation}
\chi^{+-}(mn, t - t') = <<S^{+}_{m}(t),S^{-}_{n}(t')>> = -i\theta(t
- t')<[ S^{+}_{m}(t),S^{-}_{n}(t')]_{-}>
\end{equation}
In order to
evaluate the GFs (106) and (107) we need to use the suitable
information about a ground state of the system. For the 2D spin 1/2
quantum antiferromagnet in a square lattice the calculation of the
exact ground state is a very difficult problem. In this paper we assume
the two-sublattice Neel ground state. To justify this choice one can
suppose that there are well developed short-range order
(c.f.Ref.~\cite{trap46}) or there are weak interlayer exchange
interaction which stabilize this antiferromagnetic order. According to
Neel model, the spin Hamiltonian (102) may be expressed
as~\cite{kuz17},\cite{kuz29} \begin{equation} H_{J} = \sum_{<mn>}
\sum_{\alpha,\beta}J^{\alpha \beta}{\vec S_{m\alpha}} {\vec S_{n\beta}}
\end{equation} Here $(\alpha, \beta) = (a,b)$ are the sublattice
indices.\\ To calculate the electronic states induced by hole-doping in
the spin- fermion model approach we need to calculate the energies of a
hole introduced in the Neel antiferromagnet. To be consistent with
(108) and (90) we define the single-particle fermion GF as
\begin{equation} G(k\sigma, \omega) = \pmatrix{ <<a_{a}(k\sigma) \vert
a^{+}_{a}(k\sigma)>> & <<a_{a}(k\sigma)\vert a^{+}_{b}(k\sigma)>>\cr
<<a_{b}(k\sigma)\vert a^{+}_{a}(k\sigma)>> & <<a_{b}(k\sigma)\vert
a^{+}_{b}(k\sigma)>>\cr} \end{equation} Note, that the same fermion
operators $a_{\alpha}(i\sigma)$, annihilates a fermion with spin
$\sigma$ on the $(\alpha)$-sublattice in the i-th unit cell has been
used in paper ~\cite{fur8}. The equation of motion for
the elements of GF (109) are written as \begin{equation}
\sum_{\gamma}(\omega \delta_{\alpha \gamma} - \epsilon^{\alpha
\beta}(k)) <<a_{\gamma}(k\sigma) \vert a^{+}_{\beta}(k\sigma)>> =
\delta_{\alpha \beta} - <<A(k\sigma,\alpha) \vert a^{+}_{\beta}>>
\end{equation} where \begin{equation} A(k\sigma,\alpha) =
N^{-1/2}\sum_{p}K(p)(S^{-\sigma}_{-p\alpha} a_{\alpha}(k+p -\sigma) +
z_{\sigma}S^{z}_{-p\alpha}a_{\alpha}(k+p\sigma))
\end{equation}
We make use of the IGF
approach (see Section 2)
to threat the equation of motion
(110). It may be shown that equation (110) can be
rewritten as the Dyson equation (10)
\begin{equation}
G(k\sigma,\omega) = G_{0}(k\sigma,\omega) +
G_{0}(k\sigma,\omega)M(k\sigma,\omega)G(k\sigma,\omega)
\end{equation}
Here $G_{0}(k\sigma,\omega) = \Omega^{-1}$ describes the behaviour of
the electronic subsystem in the Generalized Mean-Field(GMF)
approximation . The $\Omega$ matrix have the form
\begin{equation}
\Omega(k\sigma,\omega) =\pmatrix{
(\omega -\epsilon_{a}(k\sigma))&-\epsilon^{ab}(k)\cr
-\epsilon^{ba}(k)&(\omega -\epsilon_{b}(k\sigma))\cr}
\end{equation}
where
\begin{equation}
\epsilon_{\alpha}(k\sigma) = \epsilon^{\alpha \alpha}(k) - z_{\sigma}
N^{-1/2}\sum_{p}K(p)<S^{z}_{p\alpha}>\delta_{p,0} =
\epsilon^{\alpha \alpha}(k) - z_{\sigma}KS_{z}
\end{equation}
$$S_{z} = N^{-1/2}<S^{z}_{0\alpha}>$$
is the renormalized band energy of the holes.\\
The elements of the matrix GF $G_{0}(k\sigma,\omega)$ are found to be
\begin{equation}
G^{aa}_{0}(k\sigma,\omega) = \frac{u^{2}(k\sigma)}{\omega -\epsilon_{+}
(k\sigma)} + \frac{v^{2}(k\sigma)}{\omega - \epsilon_{-}(k\sigma)}
\end{equation}
\begin{equation}
G^{ab}_{0}(k\sigma,\omega) = \frac{u(k\sigma)v(k\sigma)}{\omega -
\epsilon_{+}(k\sigma)} - \frac{u(k\sigma)v(k\sigma)}{\omega -
\epsilon_{-}(k\sigma)} = G^{ba}_{0}(k\sigma,\omega)
\end{equation}
\begin{equation}
G^{bb}_{0}(k\sigma,\omega) = \frac{v^{2}(k\sigma)}{\omega -
\epsilon_{+}(k\sigma)} + \frac{u^{2}(k\sigma)}{\omega -
\epsilon_{-}(k\sigma)}
\end{equation}
where
\begin{equation}
u^{2}(k\sigma) = 1/2(1 - z_{\sigma}\frac{KS_{z}}{R(k)});\\
v^{2}(k\sigma) =1/2(1 + z_{\sigma}\frac{KS_{z}}{R(k)});
\end{equation}
\begin{equation}
\epsilon_{\pm}(k\sigma) = \pm R(k) = ((\epsilon^{ab}(k)^{2} +
K^{2}S^{2}_{z})^{1/2}
\end{equation}
the simplest assumption is that each sublattice is s.c. and
$\epsilon^{ \alpha \alpha}(k) = 0 (\alpha = a,b)$. In spite that
we have worked in the GFs formalism, our expressions (115) -(117)
are in accordance with the results of the Bogoliubov
(u,v)-transformation for fermions, but, of course, the present
derivation is more general. \\ The mass operator M in Dyson
equation (112), which describes hole-magnon scattering processes,
is given by as a "proper" part  of the irreducible matrix GF of
higher order
\begin{equation}
M(k\sigma,\omega) = \pmatrix{
^{(ir)}<<A(k\sigma,a)\vert A^{+}(k\sigma,a)>>^{(ir)} & ^{(ir)}<<
A(k\sigma,a)\vert A^{+}(k\sigma,b)>>^{(ir)}\cr
^{(ir)}<<A(k\sigma,b)\vert A^{+}(k\sigma,a)>>^{(ir)} &
^{(ir)}<<A(k\sigma,b)\vert A^{+}(k\sigma,b)>>^{(ir)}\cr}
\end{equation}
To find the renormalization of the spectra $\epsilon_{\pm}(k\sigma)$
and the damping of the quasiparticles it is necessary to determine the
self-energy for each type of excitations.
From the formal solution (11) one immediately obtain
\begin{equation}
G_{\pm}(k\sigma) = (\omega - \epsilon_{\pm}(k\sigma)
-\Sigma^{\pm}(k\sigma,\omega))^{-1}
\end{equation}
Here the self-energy operator is given by
\begin{equation}
\Sigma^{\pm}(k\sigma,\omega) = A^{\pm}M^{aa} \pm A_{1}(M^{ab} + M^{ba})
+ A^{\mp}M^{bb} \end{equation} where $$ A^{\pm} ={u^{2}(k\sigma)\choose
v^{2}(k\sigma)}$$ $$A_{1} = u(k\sigma)v(k\sigma)$$ Equations (121)
determines the quasiparticle spectrum with damping $(\omega = E -
i\Gamma)$ for the hole in the AFM background. Contrary to the
calculations of the hole GF in Section 4.3, the self-energy (122) is
proportional to $K^{2}$ but not $t^{2}$ (c.f.eqn.  (97))
\begin{eqnarray} M^{\alpha \beta}(k\sigma, \omega) =
N^{-1}K^{2}\sum_{q} \int_{-\infty}^{+\infty}d\omega_{1}d\omega_{2}
\frac {1 + N(\omega_{1}) - f(\omega_{2})}{\omega - \omega_{1} -
\omega_{2}}\\
\nonumber
(F^{\sigma,-\sigma}_{\alpha \beta}(q,\omega_{1})g_{\alpha
\beta}(k+q -\sigma, \omega_{2}) + F^{zz}_{\alpha \beta}(q,\omega_{1})
g_{\alpha \beta}(k+q,\omega_{2}))
\end{eqnarray}
Here functions $N(\omega)$ and $f(\omega)$ are Bose and Fermi
distributions, respectively, and the following notations have been used
for spectral intensities
\begin{eqnarray}
F^{ij}_{\alpha \beta}(q,\omega) = -\frac {1}{\pi}Im<<S^{i}_{q\alpha}
\vert S^{j}_{-q\beta}>>_{\omega}\\
\nonumber
g_{\alpha \beta}(k\sigma,\omega) =
-\frac{1}{\pi}Im<<a_{\alpha}(k\sigma)\vert
a^{+}_{\beta}(k\sigma)>>_{\omega}; \quad i,j =(+,-,z).
\end{eqnarray}
The equations (123) and (112) forms the self-consistent set of
equations for the determining of the GF (109).  Coupled equations
(123) and (112) can be solved analytically by suitable iteration
procedure. In principle, we can use, in the right-hand side of
(123) any workable first iteration step for of the relevant GFs
and find a solution by repeated iteration.  \subsection{Dynamics
of Spin Subsystem} It will be useful to discuss briefly the
dynamics of spin subsystem of the Kondo-Heisenberg model. When
calculating the spin wave spectrum of this model we shall use the
approach of Ref.~\cite{kuz17} where the quasiparticle dynamics of
the two-sublattice Heisenberg antiferromagnet has been studied
within IGF method. The contribution of the conduction electrons to
the energy and damping of the acoustic magnons in the
antiferromagnetic semiconductors within IGF scheme have been
considered in Refs.~\cite{kuz29}, \cite{kuz30}.  The main
advantage of the approach of paper~\cite{kuz17} was the using of
concept of "anomalous averages" (c.f.~\cite{kuz23}) fixing the
relevant (Neel) vacuum and providing a possibility to determine
properly the generalized mean fields. The functional structure of
required GF has the following matrix form
\begin{equation} \pmatrix{ <<S^{+}_{ka}\vert S^{-}_{ka}>> &
<<S^{+}_{ka}\vert S^{-}_{-kb}>> \cr <<S^{+}_{kb} \vert
S^{-}_{-ka}>> & <<S^{+}_{kb}\vert S^{-}_{-kb}>> \cr} = \hat
\chi(k;\omega) \end{equation}
Here the spin operators $S^{\pm}_{ka(b)} $ refer to the two sublattices
$(a,b)$.\\
The equation of motion for GF (125) after introducing the irreducible
parts  has the form~\cite{kuz30}
\begin{eqnarray}
\sum_{\gamma}((\omega +
\omega^{\alpha}_{0})\delta_{\alpha\gamma} -
\omega^{\gamma\alpha}_{k}(1 - \delta_{\alpha\gamma}))<<S^{+}_{k\gamma}
\vert B >>_{\omega} + \nonumber \\ {K \over
N^{1/2}}<S^{z}_{\alpha}><<\sigma^{+}_{k} \vert B>>_{\omega} =
<[S^{+}_{k\alpha}, B]> + << C^{ir}_{k\alpha} \vert B>>_{\omega}
\end{eqnarray}
where the following notations have been used
$$ B = {S^{-}_{-ka}\brace S^{-}_{-kb}}, \quad \alpha =(a,b)$$\\
\begin{eqnarray}
\omega^{a}_{0} =
2(<S^{z}_{b}>J_{0}
+ N^{-1/2}\sum_{q}J_{q}A^{ab}_{q}) = - \omega^{b}_{0}~; \nonumber \\
\omega^{ba}_{k} =
2(<S^{z}_{b}>J_{k}
+ N^{-1/2}\sum_{q}J_{k-q}A^{ba}_{q}) = - \omega^{ab}_{k}~; \nonumber \\
A^{ab}_{q} = \frac { 2<(S^{z}_{-qa})^{ir} (S^{z}_{qb})^{ir}>}{
2N^{1/2}<S^{z}_{a}>}
\end{eqnarray}
The  construction of the irreducible GF $<<C^{ir}_{k\alpha} \vert B>>$
is related with the operators
\begin{eqnarray}
C^{ir}_{k\alpha} = A^{ir}_{k\alpha} + B^{ir}_{k\alpha}~: \nonumber \\
A^{ir}_{ka} = {2 \over N^{1/2}} \sum_{q} J_{q}(
S^{+}_{qb}(S^{z}_{k-qa})^{ir} - S^{+}_{k-qa}(S^{z}_{qb})^{ir})^{ir}~;
\nonumber \\
B^{ir}_{ka} =
- {K \over N^{1/2}} \sum_{pq}
(S^{z}_{k-qa})^{ir}a^{+}_{p\uparrow}a_{p+q\downarrow} +
{K \over 2N} \sum_{pq\sigma}
z_{\sigma}S^{+}_{k-qa}( a^{+}_{p\sigma}a_{p+q\sigma})^{ir}
\end{eqnarray}
With the aid of (24) and (25)
the equation of motion for the mixed GF can be written as
\begin{eqnarray}
<<\sigma^{+}_{k} \vert B>> ={KN^{1/2} \over 2} \chi^{df}_{0}(k,\omega)
\sum_{\gamma} <<S^{+}_{k\gamma} \vert B>> + \nonumber \\ {K \over
2N^{1/2}} \sum_{p} {1 \over \omega_{p,k}} <<(D^{\gamma}_{pk})^{ir}
\vert B>> \end{eqnarray} Combining the equations of motion (126) and
(129) we find \begin{equation} \hat \Omega_{s} \hat \chi(k\omega) =
\hat I + \hat D_{1} \end{equation} where \begin{eqnarray} \hat
\Omega_{s} = \pmatrix{ \omega + \omega_{0} + {K^{2}S_{z} \over
2}\chi^{df}_{0} & \gamma_{2}(k)\omega_{0} + {K^{2}S_{z} \over
2}\chi^{df}_{0} \cr -(\gamma_{2}(k)\omega_{0} + {K^{2}S_{z} \over
2}\chi^{df}_{0})& \omega - \omega_{0} - {K^{2}S_{z} \over
2}\chi^{df}_{0})\cr} \nonumber \\ \hat I = \pmatrix{ 2S_{z} & 0\cr
0&-2S_{z}\cr} \end{eqnarray}
Then equation (130) can be transformed exactly into the Dyson equation
for the spin subsystem \begin{equation} \hat \chi(k\omega) = \hat
\chi^{0}(k\omega) + \hat \chi^{0}(k\omega) \hat M^{s}(k\omega) \hat
\chi(k\omega) \end{equation}
Here
\begin{equation}
\hat \chi^{0}(k\omega) = \hat \Omega_{s}^{-1}\hat I
\end{equation}
The mass operator of the spin excitations is given by the expression
\begin{equation}
\hat M^{s}(k\omega) = {1 \over 4S^{2}_{z}}
\pmatrix{ <<C^{ir}_{ka} \vert  (C^{+}_{ka})^{ir}>> &
<<C^{ir}_{ka} \vert  (C^{+}_{kb})^{ir}>> \cr <<C_{kb}^{ir} \vert
( C^{+}_{ka})^{ir}>> & <<C_{kb}^{ir} \vert ( C^{+}_{kb})^{ir}>> \cr}
\end{equation}
We are interesting here in the calculation of the spin excitation
spectrum in the generalized mean field approximation.This spectrum is
given by the poles of the GF $\hat \chi^{0}$
\begin{equation}
det \Omega_{s}(k\omega) = 0
\end{equation}
Depending of the interrelation of the parameters this spectra have
different forms. For the standard condition $2t \gg KS_{z}$ we obtain
for the magnon energy~\cite{kuz30} \begin{equation} \omega^{\pm}_{k} =
\pm \omega_{k} = \pm \Bigl ( \omega_{0}\sqrt {1 - \gamma_{2}(k)^{2}} \mp
{K^{2}S_{z} \over 2}\chi^{df}_{0}(k,\omega_{k}) \sqrt { \frac {1 -
\gamma_{2}(k)}{1 + \gamma_{2}(k)}} \Bigr ) \end{equation}
The acoustic magnon dispersion law for the $k \rightarrow 0$ is given
by \begin{equation}
\omega^{\pm}_{k} = \pm \tilde D(T) \vert \vec k \vert
\end{equation}
where the stiffness constant~\cite{kuz30}
\begin{equation}
\tilde D(T) = zJS_{z} \Bigl ( 1 - \frac {1}{\sqrt {N}S_{z}}
\sum_{q} \gamma_{2}(\vec q) A^{ab}_{q} \Bigr ) - \frac
{K^{2}S_{z}}{4N} \lim_{k \rightarrow 0}
\chi^{df}_{0}(k,\omega_{k}) \end{equation} The detailed
consideration of the spin quasiparticle damping will be done in
separate publication. Here we now proceed with calculating the
damping of the hole quasiparticles. \subsection{Damping of Hole
Quasiparticles} It is most convenient to choose as the first
iteration step in (123) the simplest two-pole expressions,
corresponding to the GF structure for a mean field, in the
following form \begin{equation} g_{\alpha \beta}(k\sigma,\omega) =
Z_{+}\delta(\omega - t_{+}(k\sigma)) + Z_{-}\delta(\omega -
t_{-}(k\sigma))
\end{equation} where $Z_{\pm}$ are the certain coefficients depending
on $u(k\sigma)$ and $v(k\sigma)$.  The magnetic excitation
spectrum corresponds to the frequency poles of the GFs  (107).
Using the results of Section 4.5 on spin dynamics of the present
model, we shall content ourselves here with the simplest initial
approximation for the spin GF occurring in (123) \begin{equation}
\frac {1}{2z_{\sigma}S_{z}}F^{\sigma -\sigma}_{\alpha
\beta}(q,\omega) = L_{+}\delta( \omega -z_{\sigma}\omega_{q}) -
L_{-}\delta(\omega + z_{\sigma}\omega_{q}) \end{equation} Here
$\omega_{q}$ is the energy of the antiferromagnetic magnons (136)
and $L_{\pm}$ are the certain coefficients (see~\cite{kuz17}).  We
are now in a position to find an explicit solution of coupled
equations obtained so far. This is achieved by using (139) and
(140) in the right- hand-side of (123). Then the hole self-energy
in 2D quantum antiferromagnet for the low-energy quasiparticle
band $t_{-}(k\sigma)$ is \begin{eqnarray}
\Sigma^{-}(k\sigma,\omega) = \frac {K^{2}S_{z}}{2N}
\sum_{q}Y_{1}^{2} (\frac {1 + N(\omega_{q}) -
f(t_{-}(k-q))}{\omega - \omega_{q} - t_{-}(k-q)} + \frac{N(\omega)
+ f(t_{-}(k+q))}{\omega + \omega_{q} - t_{-}(k+q)})\\ \nonumber +
\frac{2K^{2}S^{2}_{z}}{N} \sum_{qp}Y_{2}^{2} \frac{
N(\omega_{q+p})(1 + N(\omega_{q})) + f(t_{-}(k+p))(N(\omega_{q}) -
N(\omega_{q+p}))}{\omega + \omega_{q+p} - \omega_{q} - t_{-}(k+p)}
\end{eqnarray} Here we have used the notations
$$Y_{1}^{2} = (U_{q} + V_{q})^{2};\quad Y_{2}^{2} =
(U_{q}U_{q+p} - V_{q}V_{q+p})^{2}$$ where the coefficients $U_{q}$ and
$V_{q}$ appears as a results of explicit calculation of the mean-field
magnon GF ~\cite{kuz17}, \cite{kuz30}.\\ A  remarkable feature of
this result is that our expression (141) accounts for the hole-magnon
inelastic scattering processes with the participation of one or two
magnons. \\ The self-energy representation
in a self-consistent form (123) provide a possibility to model the
relevant spin dynamics by selecting spin-diagonal or spin-off-diagonal
coupling as a dominating or having different characteristic frequency
scales. As a workable pattern, we consider now the static trial
approximation for the correlation functions of the magnon
subsystem~\cite{kuz17} in the expression (123).  Then the following
expression is readily obtained \begin{eqnarray} M^{s}_{\alpha
\beta}(k\sigma,\omega) = \frac {K^{2}}{N}\sum_{q}\int_{
-\infty}^{+\infty} \frac{d\omega'}{\omega -
\omega'}(<S^{-\sigma}_{-q\beta}S^{\sigma}_{q\alpha}>g_{\alpha \beta}(
k+q -\sigma,\omega')\\
\nonumber
+ <(S^{z}_{-q\beta})^{ir}(S^{z}_{q\alpha})^{ir}>g_{\alpha
\beta}(k+q\sigma,\omega'))
\end{eqnarray}
Taking into account (141) we find the following approximate form
\begin{equation}
\Sigma^{-}(k\sigma, \omega) \approx \frac{K^{2}}{2N}\sum_{q}\frac{
\chi^{-+}(q) + \chi^{z,z}(q)}{\omega - t_{-}(k+q)}(1 - \gamma_{3}(q))
\end{equation}
It should be noted, however, that in order to make this kind of study
valuable as one of the directions to studying the mechanism of HTSC the
binding of quasiparticles must be taking into account. This very
important problem deserves the
separate consideration. In spite of formal analogy of the our model (99)
with that of a Kondo lattice, the physics are different.
There is a dense system of spins interacting with a smaller
concentration of holes.
This question is in close relation with the right definition of the
magnon vacuum for the case when $K \not= 0$.\\ In this Section we has
considered the simplest possibility, assuming that dispersion relation
$\epsilon^{\alpha \alpha}(k) = 0$ ($\alpha = a,b$).  In
paper~\cite{dag38} a model of hole carriers in an antiferromagnetic
background has been discussed, which explains many specific properties
of cuprates. The effect of strong correlations is contained in the
dispersion relation of the holes. The main assumption is that the
influence of antiferromagnetism and strong correlations is contained in
the special dispersion relation, $\epsilon(k)$, which was obtained
using a numerical method. The best fit corresponds to\cite{dag38}
\begin{equation}
\epsilon(k) = -1.255 + 0.34\cos k_{x}\cos k_{y} + 0.13(\cos2k_{x} +\cos2k_{y})
\end{equation}
As a result, the main effective contribution to $\epsilon(k)$ arises from
hole hopping between sites belonging to the same sublattice, to avoid
distorting the antiferromagnetic background.\\  Our
IGF method is essentially self-consistent, i.e. do not depends on the
special initial form for the hole propagator. For the self-consistent
calculation by iteration of the self-energy (123) we can take as the
fist iteration step the expression (139) with the dispersion relation
(144) or another suitable form. This must be done for the calculation
mean-field GF (113) and dispersion relation (119) too. \\
To summarize, in Section 4 we have presented
calculations for normal phase of HTSC, which are describable in terms
of the spin-fermion model. We have characterized the true quasiparticle
nature of the carriers  and the role of magnetic correlations. It was
shown that the physics of spin-fermion model can be understood in terms
of competition between antiferromagnetic order on the $CuO_{2}$-plane
preferred by superexchange J and the itinerant motion of carriers.
In the light of this situation it is clearly of interest to explore
in details how the hole motion influence the antiferromagnetic
background. Considering that the carrier-doping
results in the HTSC for the realistic parameters range $t\gg J,K$,
corresponding the situation in oxide superconductors, the careful
examination of the collective behaviour of the carriers for a
moderately doped system must be performed. It seems that this behaviour
can be very different from that of single hole case. The problem of the
coexistence of the suitable Fermi-surface of mobile fermions and the
antiferromagnetic long range or short range order has to be clarified.
\section{Conclusions} We have been concerned in
this paper with establishing  the essence of quasiparticle
excitations of charge and spin degrees of freedom within a
generalized spin-fermion model, rather than with their detailed
properties. We have considered the generalized $d-f$ model and
Kondo-Heisenberg model as the most typical examples but the
similar calculation can be performed for other analogous models.
To summarize, we therefore reanalyzed within IGF approach the
quasiparticle many body dynamics of the generalized spin-fermion
model in a way which provides us with an effective and workable
scheme for consideration of the quasiparticle spectra and their
damping for the correlated systems with complex spectra. The
calculated temperature behaviour of the damping of acoustical
magnons (67) can be useful for analysis of the experemental
results for heavy rare-earth metals like $Gd$ ~\cite{coq1}. The
present analysis of the 2D Kondo-Heisenberg model complements the
previous analytical~\cite{shr11} and numerical~\cite{pre35}
studies, showing clearly the important role of the damping
effects. \\ We have considered from a general point of view the
family of solutions for itinerant lattice fermions and localized
spins on a lattice, identifing the type of ordered states and then
derived explicitly the functional of generalized mean fields and
the self-consistent set of equations which describe the
quasiparticle spectra and their damping in the most general way.
While such generality is not so obvious in all applications, it is
highly desirable in treatments of such complicated problems as the
competition and interplay of antiferromagnetism and
superconductivity, heavy fermions and antiferromagnetism etc.,
because of the non-trivial character of coupled equations which
occur there. The problem of the coexistence of HF and magnetism is
extremely nontrivial~\cite{cole},\cite{coq6} many-body problem and
have no appropriate solution in spite that there are many
experimental evidences of the competition and interplay of HF and
antiferromagnetism~\cite{coq6}. Both these problems are subject of
current but independent research.\\ Another development of the
present approach is the consideration of the competition and
interplay of itinerant and localized magnetism and
antiferromagnetism of the doped manganite perovskites where the
interrelation between parameters of the spin-fermion model is
quite different and the new scheme of approximation should be
invented.  Especially, the situation , when Hund rule interaction
is very large but finite should be carefully analyzed.  It would
be interesting to understand on a deeper level the relationship
between different possible phase states in manganates and various
ordered magnetic states within the generalized spin-fermion
model.\\ In conclusion, we have demonstrated that the Irreducible
Green's Functions approach is a workable and efficient scheme for
the consistent description of the quasiparticle dynamics of
complicated many body models.


\begin{thebibliography} {8}
\bibitem{coq1} B. Coqblin
{\em The Electronic Structure of Rare-Earth Metals and Alloys: the
Magnetic Heavy Rare-Earths} (Academic Press, N.Y., London, 1977).
\bibitem{don2} S.Doniach in  {\em Theory of Magnetism in
Transition Metals }, E.Fermi School, (Varenna,
1966) p.255.
\bibitem{kuz3} A. L. Kuzemsky {\em et al} in
{\em Crystalline Electric Field Effects in f-Electrons Magnetism},
edited by T. R. Guertin {\em et al}, (Plenum Press, N.Y., 1982),
p.219.
\bibitem{imad}
M.Imada, A. Fujimori and Y. Tokura, {\em Rev.Mod.Phys.} {\bf 70},
1039 (1998).
\bibitem{kuz4}
D. Marvakov, J. Vlahov and A. L. Kuzemsky, {\em J.Physics C:Solid
State Phys.}{\bf 18}, 2871 (1985).
\bibitem{kuz5} D.Marvakov,
A. L. Kuzemsky and J. Vlahov, {\em Physica} {\bf B138}, 129
(1986).
\bibitem{cole}
P. Coleman and N. Andrei, {\em J.Physics:Condens.Matter }{\bf 1},
4057 (1989).
\bibitem{coq6} B.Coqblin, J.R.Iglesias and C.Lacroix, {\em JMMM }
{\bf 177-181}, 433 (1998).
\bibitem{pre7}
P. Prelovsek, {\em Phys.Lett.} {\bf A126}, 287 (1988).
\bibitem{fur8}
N. Furukawa and M. Imada, {\em J.Phys.Soc.Jpn.} {\bf 59}, 1771
(1990).
\bibitem{ino9}
J. Inoue and S. Maekawa, {\em J.Phys.Soc.Jpn.} {\bf 59}, 2110
(1990).
\bibitem{kuz10} A. L. Kuzemsky, in:  {\em Superconductivity and Strongly
Correlated Electron Systems}, eds.C. Noce et al (World Scientific,
Singapoure, 1994) p.346.
\bibitem{shr11}
J. R. Schrieffer, {\em J.Low Temp.Phys.} {\bf 99}, 397 (1995).
\bibitem{kuz12}
A. L. Kuzemsky and D. Marvakov, {\em Modern Phys.Lett.} {\bf B9},
1719 (1995).
\bibitem{dege13} P. de Gennes, {\em Phys.Rev.} {\bf 118},
141 (1960).
\bibitem{mae14} S. Ishihara, J. Inoue and S. Maekawa, {\em
Phys.Rev.} {\bf B55}, 8280 (1997).
\bibitem{kuz15} A. L. Kuzemsky,
{\em Theor.  Math. Phys.} {\bf 36}, 208 (1978).
\bibitem{kuz16}
A. L. Kuzemsky, {\em Doklady Acad.Nauk SSSR} {\bf 309}, 323
(1989).
\bibitem{kuz17} A. L. Kuzemsky and D. Marvakov, {\em Theor. Math. Phys.}
{\bf 83}, 147 (1990).
\bibitem{kuz18} A. L. Kuzemsky, {\em Phys.Lett.} {\bf
A 153}, 446 (1991).
\bibitem{kuz19} A. L. Kuzemsky, J.-C. Parlebas
and H. Beck, {\em Physica} {\bf A 198}, 606 (1993).
\bibitem{kuz20}
A. L. Kuzemsky, {\em Nuovo Cimento} {\bf B 109}, 829 (1994).
\bibitem{kuz21} A. L. Kuzemsky, {\em Intern.J.Modern Phys.}  {\bf B 10},
1895 (1996).
\bibitem{kuz22} A. L. Kuzemsky, {\em Molecular Phys.Rep.}
{\bf 17}, 221 (1997).
\bibitem{kuz23} A. L. Kuzemsky,  {\em Physica}
{\bf A 267}, 131 (1999).
\bibitem{tyab24} S. V. Tyablicov, {\em
Methods in the Quantum Theory of Magnetism} (Plenum Press, New York,
1967).
\bibitem{kuz25}
A. L. Kuzemsky, in {\it Physics of Elementary Particles and Atomic
Nuclei}, ed. by N.N.Bogolubov et al (Atomisdat, Moscow, 1981),
{\bf 12}, p.366.
\bibitem{kuz26} A. L. Kuzemsky, A. Holas and N. M. Plakida, {\em Physica}
{\bf B 122}, 168 (1983).
\bibitem{kuz27} A. L. Kuzemsky and
A. P. Zhernov, {\em Intern.J.Modern Phys.} {\bf B4}, 1395 (1990).
\bibitem{plak}
N. M. Plakida, {\em Phys.Lett.} {\bf 43A}, 481 (1973).
\bibitem{hub28} J. Hubbard,
{\em Proc.Roy.Soc.} {\bf A 276} (1963) 238.
\bibitem{kuz29}
D. Marvakov, R. Ahed and A. L. Kuzemsky, {\it Bulgarian J.Phys.}
{\bf 17}, 191 (1990).
\bibitem{kuz30}
D. Marvakov, R. Ahed and A. L. Kuzemsky, {\it Bulgarian J.Phys.}
{\bf 18}, 8 (1991).
\bibitem{kane31} C. L. Kane, P. A. Lee and N. Read, {\em Phys.Rev.} {\bf
B39}, 6880 (1989).
\bibitem{su32} Z. B. Su, Y. M. Li, W. J. Lai and L. Yu.
{\em Phys.Rev.Lett.} {\bf 63}, 1318 (1989).
\bibitem{su33} Z. B. Su, Y. M. Li, W. J. Lai and L. Yu.
{\em Intern.J.Modern Phys.}
{\bf B3}, 1913 (1989).
\bibitem{hu34} J. H. Hu, C. S. Ting and T. K. Lee,
{\em Phys.Rev.}  {\bf B43}, 8733 (1990).
\bibitem{pre35} A. Ramsak and P. Prelovsek,{\em
Helvetica Physica Acta }, {\bf 65}, 446 (1992).
\bibitem{su36} Y. M. Li, W. J. Lai,
Z. B. Su and L. Yu, {\em Phys.Rev.} {\bf B47}, 7938 (1993).
\bibitem{su37} L. Yu, Z. B. Su and Y. M. Li,
{\em Chinese J.Phys.} {\bf 31}, 579 (1993).
\bibitem{dag38}
E. Dagotto, A. Nazarenko and M. Boninsegni, {\em Phys.Rev.Lett.}
{\bf 73}, 728 (1994).
\bibitem{ino39}
J. Inoue, K. Iiyama, S. Kobayashi and S. Maekawa, {\em Phys.Rev. }
{\bf B49}, 6213 (1994).
\bibitem{weng40} Z. Y. Weng, N. D. Sheng and C. S. Ting,
{\em Phys.Rev.}  {\bf B52}, 637 (1995).
\bibitem{gan41} J. Gan and Per Hedegard,
{\em Phys.Rev.}  {\bf B53}, 911 (1996).
\bibitem{dot42} A. V. Dotsenko,
{\em Phys.Rev.}  {\bf B57}, 6917 (1998).
\bibitem{yang43}
J. Yang, W. G. Yin and C. D. Gong, {\em Modern Phys.Lett.} {\bf
B12}, 205 (1998).
\bibitem{feng44}
Shiping Feng, {\em
cond-mat/9809098}.
\bibitem{emer45}
V. J. Emery, S. A. Kivelson and O. Zachar, {\em cond-mat/9810155}.
\bibitem{trap46} U. Trapper, D. Ihle and H. Fehske,
{\em Intern.J.Modern Phys.}
{\bf B11}, 1337 (1997).
%
\end{thebibliography}
 \end{document}